\documentclass[fleqn,usenatbib,useAMS]{mnras}

\usepackage[T1]{fontenc}

\DeclareRobustCommand{\VAN}[3]{#2}
\let\VANthebibliography\thebibliography
\def\thebibliography{\DeclareRobustCommand{\VAN}[3]{##3}\VANthebibliography}

\usepackage{ae,aecompl}
\usepackage{newtxtext,newtxmath}

\usepackage{graphicx}	
\usepackage{amsmath}	
\usepackage{multicol}        
\usepackage{bm}		
\usepackage{pdflscape}	
\usepackage{longtable}  
\usepackage{caption}
\usepackage[flushleft]{threeparttable}
\usepackage{multirow}   
\usepackage{xfrac}   
\usepackage{xcolor} 

\newcommand{\kms}{\,km\,s$^{-1}$} 
\newcommand{\tsky}{\,$T_{\text{sky}}$} 
\newcommand{\uJy}{\,$\upmu$Jy} 
\newcommand{\smin}{\,$S_{\text{min}}$} 
\newcommand{\dm}{~pc~cm$^{-3}$} 
\newcommand{\lum}{mJy~kpc$^{2}$} 
\newcommand{\gray}{$\upgamma$-ray}
\newcommand{\grays}{$\upgamma$-rays}
\newcommand{\edot}{$\dot{E}$}

\def\psrchive{\mbox{\textsc{psrchive}}}
\def\pulsarx{\mbox{\textsc{pulsar}X}}
\def\mosaic{\mbox{\textsc{mosaic}}}
\def\clfd{\mbox{\textsc{clfd}}}
\def\seekat{\mbox{S\textsc{ee}KAT}}
\def\filtool{\mbox{\texttt{filtool}}}
\def\dspsr{\mbox{\texttt{DSPSR}}}
\def\pdmp{\mbox{\texttt{pdmp}}}
\def\psrA{\mbox{PSR J1831$-$0941}}
\def\psrB{\mbox{PSR J1818$-$1502}}

\renewcommand{\dotfill}{%
  \leavevmode\cleaders\hbox to 1.00em{\hss .\hss }\hfill\kern0pt }
\title[TRAPUM search for Pulsars in Supernova Remnants I]{TRAPUM search for pulsars in supernova remnants and pulsar wind nebulae $-$ I. Survey description and initial discoveries}

\author[J. D. Turner et al.]{J. D. Turner$^{1}$\thanks{E-mail: \href{mailto:james.turner-13@postgrad.manchester.ac.uk}{james.turner-13@postgrad.manchester.ac.uk}}, B. W. Stappers$^{1}$, E. Carli$^{1}$, E. D. Barr$^{2}$, W. Becker$^{2,3}$, J. Behrend$^{2}$, R. P. Breton$^{1}$, \newauthor S. Buchner$^{4}$, M. Burgay$^{5}$, D. J. Champion$^{2}$, W. Chen$^{2}$, C. J. Clark$^{6,7}$, D. M. Horn$^{4}$, E. F. Keane$^{8}$, \newauthor M. Kramer$^{2}$, L. Künkel$^{9}$, L. Levin$^{1}$, Y. P. Men$^{2}$, P. V. Padmanabh$^{6,7,2}$, A. Ridolfi$^{2,5}$, \newauthor V. Venkatraman Krishnan$^{2}$ \\
$^{1}$Jodrell Bank Centre for Astrophysics, Department of Physics and Astronomy, The University of Manchester, Manchester M13 9PL, UK\\
$^{2}$Max-Planck-Institut für Radioastronomie, Auf dem Hügel 69, D-53121 Bonn, Germany\\ $^{3}$Max-Planck-Institut für extraterrestrische Physik, Giessenbachstraße, 85748 Garching, Germany\\ $^{4}$South African Radio Astronomy Observatory (SARAO), 2 Fir Street, Black River Park, Observatory, Cape Town, 7925, South Africa \\ $^{5}$INAF - Osservatorio Astronomico di Cagliari, Via della Scienza 5, I-09047 Selargius (CA), Italy\\ $^{6}$ Max Planck Institute for Gravitational Physics (Albert Einstein Institute), D-30167 Hannover, Germany\\ $^{7}$Leibniz Universit\"{a}t Hannover, D-30167 Hannover, Germany\\ $^{8}$School of Physics, Trinity College Dublin, College Green, Dublin 2, D02 PN40, Ireland\\ $^{9}$Department of Physics and Astronomy, University of British Columbia, 6224 Agricultural Road, Vancouver, BC V6T 1Z1, Canada \\
}

\date{Accepted XXX. Received YYY; in original form ZZZ}

\pubyear{2024}

\begin{document}
\label{firstpage}
\pagerange{\pageref{firstpage}--\pageref{lastpage}}
\maketitle

\begin{abstract}
We present the description and initial results of the TRAPUM (TRAnsients And PUlsars with MeerKAT) search for pulsars associated with supernova remnants (SNRs), pulsar wind nebulae and unidentified TeV emission. The list of sources to be targeted includes a large number of well-known candidate pulsar locations but also new candidate SNRs identified using a range of criteria. 
Using the 64-dish MeerKAT radio telescope, we use an interferometric beamforming technique to tile the potential pulsar locations with coherent beams which we search for radio pulsations, above a signal-to-noise of 9, down to an average flux density upper limit of 30~\uJy. This limit is target-dependent due to the contribution of the sky and nebula to the system temperature. Coherent beams are arranged to overlap at their 50 per cent power radius, so the sensitivity to pulsars is not degraded by more than this amount, though realistically averages around 65 per cent if every location in the beam is considered. We report the discovery of two new pulsars; \psrA~is an adolescent pulsar likely to be the plerionic engine of the candidate PWN G20.0+0.0, and \psrB~appears to be an old and faint pulsar that we serendipitously discovered near the centre of a SNR already hosting a compact central object. The survey holds importance for better understanding of neutron star birth rates and the energetics of young pulsars.
\end{abstract}

\begin{keywords}
Pulsars: general -- ISM: supernova remnants
\end{keywords}



\section{Introduction}\label{intro}

Pulsars are rapidly spinning, highly magnetised neutron stars (NSs), detected if the radiation that is emitted from their magnetic poles sweeps across the line of sight of an observer. NSs are predominantly thought to be born from the core-collapse supernova explosions of fuel-exhausted early spectral type main sequence stars \citet{Baade1934}. Pulsars emit spin-powered radiation for far longer than their supernova remnants (SNRs) are visible \citep[e.g.][]{Braun1989}, therefore identifiable supernova sites offer the best prospects for uncovering young pulsars.

SNRs are the expanding front of ejecta and swept up interstellar medium (ISM) resulting from supernovae (SNe). It is now understood that about 75 per cent of SNe are Type Ib/c or Type II \citep{Cappellaro1999}, all of which are triggered by the core-collapse of stars of initial mass $\ga~9~\text{M}_{\sun}$ \citep{Heger2003} and form a NS. The remainder are Type Ia supernovae, which are the result of either a thermonuclear deflagration \citep{Nomoto1984} of an accreting white dwarf star at the Chandrasekhar mass limit of $1.4~\text{M}_{\sun}$ \citep{Chandrasekhar1931} or a collision between two white dwarfs \citep{Raskin2009, Rosswog2009}. Type Ia SNe do not produce a neutron star, therefore an effective pulsar search strategy should exclude SNRs that show evidence of a thermonuclear origin, which are primarily distinguished from core-collapsed remnants by a strong presence of iron in the optical spectra \citep[see e.g.][for a review]{Vink2020}.
It is possible that the most massive 1 per cent of massive stars undergo so-called pair-instability SNe which does not produce a neutron star but does result in a SNR \citep{Heger2003}.

\begin{figure}
    \centering
    \includegraphics[width=1.0\columnwidth]{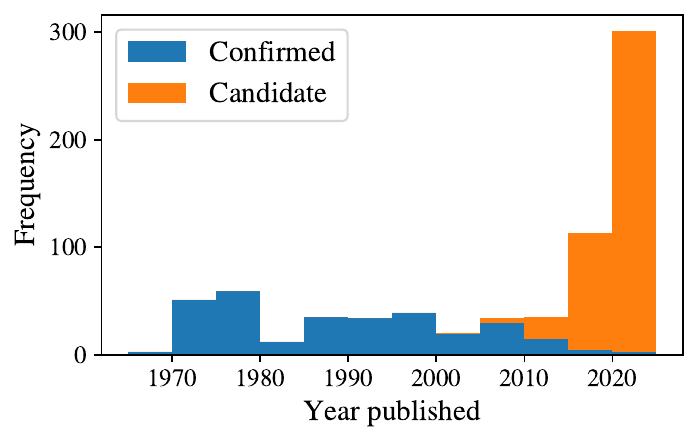}
    \caption{Histogram of new SNRs identified in the radio regime binned by their date reported in the literature. Bins are 5 years in width. The blue columns constitute the confirmed SNRs listed in G22.}
    \label{fig:pub_years}
\end{figure}

There are 303 confirmed SNRs listed in the 2022 version of the Galactic SNR Catalogue\footnote{http://www.mrao.cam.ac.uk/surveys/snrs/} \citep[][henceforth referred to as G22]{Green2022}. Radio continuum imaging surveys such as those with the Molonglo Observatory Synthesis Telescope \citep[MOST,][]{Clark1975, Gray1994, Whiteoak1996} or the Multi-Array Galactic Plane Imaging Survey \citep[MAGPIS,][]{Helfand2006, Brogan2006} with the Very Large Array (VLA) have reported regular batches of new sources. New instruments and surveys with a high sensitivity and resolution, even at lower frequencies, have identified a plethora of new candidates, particularly small diameter shells of low surface brightness on the Galactic plane. Nearly 200 of these are from three surveys; the 1-2 GHz HI, OH, Recombination line survey of the Milky Way \citep[THOR,][]{Anderson2017}, the GaLactic and Extragalactic
All-sky Murchison Widefield Array survey \citep[GLEAM,][]{Hurley-Walker2019b}, and the GLOSTAR (GLObal view of STAR formation in the Milky Way) survey \citep{Dokara2021}. The success of these surveys is demonstrated in \autoref{fig:pub_years}, which shows an accelerating discovery rate for new SNRs. Based on an all-type SN rate of one event per 40 yr \citep{Tammann1994} and a dissipation timescale of approximately 60 kyr \citep{Frail1994}, there should be around 1500 visible radio SNRs in the Galaxy. 
\autoref{fig:pub_years} includes 200 new candidates identified by MeerKAT and reported by \citet{Anderson2023}, though we do not include these unpublished targets in the survey reported here.

Most SNRs are shell-type at radio and X-ray wavelengths \citep{Green2019}, where only emission from the shock front is visible. A smaller number of shells encompass non-plerionic radio and non-thermal X-ray features of increasing brightness towards the remnant centre, and are wholly referred to as filled centre SNRs. These are not to be confused with mixed-morphology SNRs, which exhibit central soft X-ray emission in an apparent delineation from the Sedov-Taylor description of expansion \citep{Sedov1946, Taylor1950} as applied to SNR evolution, due to a dense but uniform ISM \citep{Rho1998}. If a shell encloses a plerion then they are conjointly described as a composite-type SNR. A plerion, or pulsar wind nebula (PWN), is formed when the magnetised pulsar wind is confined behind a termination shock, where the wind and pressures equalise \citep[e.g.][]{Torres2013}. The pressure confining the wind can be due to the over-dense SNR interior medium, or if the pulsar is moving supersonically. The post-shock particle radiation is dominated by synchrotron emission at radio wavelengths and X-rays, and includes \grays~from inverse Compton scattering of low energy photons \citep[see e.g.][and references therein]{Slane2017,Mitchell2022}.
Supersonic pulsar motion can be supplied by a kick from an asymmetric core collapse \citep{Podsiadlowski2005}. 
The kick velocity, which averages somewhere between 250 and 440~\kms~\citep{Hansen1997,Hobbs2005}, is occasionally sufficient for a NS to depart a still-visible SNR, e.g. the `Frying Pan'\citep{Camilo2009a}, the `Mouse'\citep{Camilo2002b} and the `Mini Mouse'\citep{Motta2023}. The bow-shock directs the energetic particles behind the pulsar, forming a tail \citep{Kargaltsev2017}.

PWNe are sources of very high energy (VHE) \grays. This emission is well described by models of energetic leptons injected by the pulsar wind \citep[see e.g.][and references therein]{Slane2017}, and can extend into the PeV regime \grays~\citep{Cao2021}. Due to the much slower cooling timescale than for pure synchrotron X-rays, the TeV component is more representative of the cumulative energy supplied by the spin-down luminosity of the pulsar over its lifetime, $\tau$, i.e. \(\int_{0}^{\tau} \)\edot\((t) \,dt\) \citep{Kargaltsev2013}, whereas the X-ray power is some fraction of the instantaneous output, \edot. The High Energy Stereoscopic System (H.E.S.S.), located in Namibia, is an array of \v{C}erenkov telescopes detecting \grays~from 0.1 to 100 TeV \citep{Bolmont2014}. A full survey of the Galactic plane has identified a total of 95 VHE sources \citep{Abdalla2018a}, the majority of which are firmly or potentially associated with known PWNe \citep{Abdalla2018b}. Sources of non-plerionic origin include hadronic VHE emission due to SN shocks colliding with molecular clouds, or from cataclysmic binary systems. According to the catalogue of TeV sources \texttt{tevcat}\footnote{\href{http://tevcat.uchicago.edu/}{http://tevcat.uchicago.edu/}}, 25 TeV sources seen by H.E.S.S. remain unidentified. Aside from PWNe, H.E.S.S. has also helped uncover an interesting new population of TeV halos, regions of VHE \grays~extending over a degree away from a pulsar. It is possible this emission is from a relic wind nebula where the energetic leptons would, in the case of a pulsar that has left the SNR shell \citep{Lopez-Coto2022}, not be bound within the magnetised wind bubble as is expected for young PWNe \citep{Gaensler2006}. Thus TeV halos could be the final stage in the life cycle of PWNe. As many as 100 halos could be detectable in the near future \citep{Sudoh2019,Martin2022}. TeV sources, therefore, are useful indicators for the presence of undetected young or adolescent pulsars.

All the sources discussed here constitute a zoo of targets that could contain many undiscovered young pulsars. The discovery rate for PSR-SNR systems has slowed in recent years; 63 out of the 303 G22 SNRs have evidence for a neutron star counterpart. It should be noted that already 20 years ago there were 51 associations \citep{Kaspi2002}. Additionally, 63/303 is consistent with the fraction of pulsars estimated to be visible from Earth assuming a beaming fraction between 1/6 to 1/5 \citep{Taylor1977,Biggs1996}. Both points suggest that a ceiling on finding new pulsars in known SNRs is approaching. Nevertheless, the jump in new SNR candidates (\autoref{fig:pub_years}), new VHE sources and the high sensitivity of current generation telescopes like MeerKAT motivate a new, large survey to discover many more young pulsars. The establishment of a more statistically significant sample is necessary for deepening our understanding of the pulsar population, the energy budgets of PWN systems, the coupling of pulsars to the evolution of their remnants, and constraining the magnitude of effects that prohibit detections of some young pulsars.

Many SNRs and PWNe are located close to, or on the Galactic plane, and therefore are covered by the footprint of many non-targeted pulsar surveys. Targeted searches are usually able to go much deeper by centring the beam pattern directly on the remnants and by integrating longer than the non-targeted survey dwell times. However, of the 40 pulsars associated with SNRs or PWNe for which radio pulsations were detected first, a minority of 12 were discovered in targeted searches while 28 were found by non-targeted surveys \citep{Manchester2005}. On several occasions, searches with sensitivities of order 1~mJy of many SNRs have produced no new discoveries; \citet{Kaspi1996} searched 39\footnote{40 are listed in the survey but one was later found to be the north-western segment of the catalogued G350.0-2.0 \citep{Green2019}} Southern SNRs using Murriyang, the 64-m radio telescope at Parkes, \citet{Gorham1996} searched 17 remnants with the 305~m Arecibo telescope, \citet{Sett2021} focused on 5 remnants with the Green Bank telescope and searched deeper than any previous attempts, but did not see anything new, and \citet{Straal2019} were fruitless in their different approach of using LOFAR to detect steep-spectrum pulsars. In a search of 33 Northern remnants with the Lovell telescope, \citet{Lorimer1998} found only one pulsar that is likely to be associated with the target \citep{Tian2006}.

It might be argued that it is preferable to focus on targeting pulsar tracers such as compact VHE and radio emission from PWNe, as these searches have often succeeded in detecting new radio pulsars \citep[e.g.][]{Wolszczan1991, Camilo2009}.
The difficulty in explaining the non-detection of a radio pulsar in any given SNR is closely linked to an incomplete understanding of selection effects. For example, a radio pulsar may escape detection if the pulses are substantially affected by the circumstellar medium and the ISM. Frequency-dependent dispersion due to the cold plasma, which delays radio waves proportionally to $\nu^{-2}$, risks smearing pulsar signals within frequency channels beyond recovery. Scattering, which is caused by a turbulent medium, can broaden a pulse and spread the signal too thinly, masking it in a similar fashion to dispersion but with a weaker frequency dependence of $\nu^{-4}$. 
Alternatively, a pulsar may also not be intrinsically luminous enough to surpass the signal-to-noise threshold. These explanations for a non-detections are not mutually exclusive, nor is the size of their effect on pulsar discovery truly constrained.


New telescopes like MeerKAT provide increased sensitivity to dimmer radio pulsars. Taking the (pseudo-) luminosity, $L_{\nu}\,=\,S_{\nu}D^{2}$, where $S_{\nu}$ is the pulsar flux density at a frequency, $\nu$, the distribution $N(L_{\nu})$ is a power law of negative index \citep{Bagchi2013}. The unavoidable bias of pulsar searches against distant and intrinsically dim sources, despite being more populous, limits modelling the behaviour of $N(L_{\nu})$ below 0.1~\lum~\citep{FaucherGiguere2006, Lorimer2006}. $N(L_{\nu})$ must turn over and fall to zero at some unknown value of $L$. MeerKAT provides the opportunity to probe the low-luminosity population, improve models of $N(L_{\nu})$ and maximise pulsar discovery.

Increasing the sample of young radio pulsars would also provide constraints on the Galactic NS birth rate. The lowest estimate by \citet{Keane2008}, amongst all the model and method-dependent formation rates they considered, is 5.7$^{+4.1}_{-2.7}$~century$^{-1}$ for all NS manifestations combined. This value exceeds the independently measured CCSN rate of 1.9$\pm$1.1~century$^{-1}$ from the Galactic content of $^{26}$Al \citep{Diehl2006}, though the uncertainties just about overlap. \citet{Keane2008} suggest two possible solutions to this inconsistency. Either there is an evolutionary link between some classes of NSs which would shrink the total birth rate, or this is possibly a case of underestimated uncertainty or overestimated rates. The most constrained birth rate between the NS types is that of the radio pulsar, from which the others are extrapolated. Therefore more informed estimates of this birth rate will reduce the uncertainty propagating through to the total.

Another manifestation of young NSs are the central compact objects (CCOs). There are ten known CCOs and a further four objects are candidates. These objects have only been seen in X-rays, which are of predominantly thermal origin and of steady flux. The rotation periods of three CCOs has been measured; PSR J1210$-$5226/1E~1207.4$-$5209 which rotates every 424 ms \citep{Zavlin2000}, PSR J1852+0040 in Kes 79 with a period of 105 ms \citep{Gotthelf2005} and PSR J0821$-$4300 with a period of 113 ms \citep{Gotthelf2009b}. This cements CCOs as neutron stars, but the characteristic age derived from assuming the rotational evolution of a magnetic dipole would be $>$100~Myr, seemingly at odds with the youth of their remnants. 
Transient X-ray emission from many CCOs can be explained as surface hot spots (see \citealt{DeLuca2017} for a review). Deep radio searches that detect radio pulses or otherwise set deep upper limits would help to understand whether CCOs are faint, quiescent or non-emitting at radio wavelengths.

This survey is a part of the TRAPUM (TRAnsients and PUlsars with MeerKAT) Large Survey Project \citep{Stappers2016}, which has collectively discovered over 200 pulsars\footnote{TRAPUM discoveries are listed at \href{http://trapum.org/}{http://trapum.org/}} using the MeerKAT telescope. Having presented the motivation for deep searches of many SNRs and PWNe with MeerKAT, we continue in Section \ref{obs} by describing the front and backend infrastructure and how the design of the survey seeks to minimise some of the aforementioned selection effects. The search pipeline is described in Section \ref{search}, the first discoveries of the survey are presented in Section \ref{results} and a discussion of these pulsars follows in Section \ref{discussion}.

\section{Observations}\label{obs}
\subsection{Beamforming with MeerKAT}\label{specs}
MeerKAT is the most sensitive radio telescope in the Southern hemisphere \citep{Camilo2018a, Jonas2018}. It is comprised of 64 dishes of Gregorian-offset design, each 13.5 m in diameter and installed with three dual polarisation receivers that collectively offer a wide frequency coverage; the L-band receiver (856-1712 MHz) \citep{Lehmensiek2012}, the Ultra High Frequency (UHF)-band receiver (544-1088 MHz) \citep{Lehmensiek2014a} and the S-band receiver (1750-3500 MHz) \citep{Barr2018}. \autoref{tab:specs} lists the specifications for our time-domain observations which so far have exclusively used the L-band receiver. The presence of narrowband radio frequency interference (RFI) sometimes causes losses of up to 20 per cent of the band.
\begin{table}
\begin{center}
\caption{A summary of specifications that are utilised for searching with the L-band receivers of MeerKAT.}\label{tab:specs}
\begin{tabular}{ll}
\hline
Parameter                                   & Value(s)      \\
\hline
Maximum number of dishes, $N_{\text{d}}$        & 64            \\
Maximum gain, $G$ (K Jy$^{-1}$) & 2.8           \\
Central frequency, $\nu$ (MHz)              & 1284          \\
Bandwidth, $\Delta\nu$ (MHz)                & 856           \\
Effective bandwidth, $\Delta\nu_{\text{eff}}$ (MHz)& 685    \\
Number of channels                          & 4096          \\
Sampling time, $t_{\text{samp}}$ ($\upmu$s) & 153.121 or 306.242    \\
Number of polarisations, $N_{\text{p}}$     & 2              \\
\hline
\end{tabular}
\end{center}
\end{table}

Voltage-based beamforming functions by coherently summing the signals arriving at each receiver, using a set of geometric delays that correct to a reference point in the array. This forms a tied-array beam, or coherent beam (CB), with a size dependent on the projected baselines of the telescopes. Up to 780 CBs\footnote{Data ingestion limits impose a maximum beam number of 780 beams for 4096 channnels and a sampling time of 306~$\upmu$s.} can be packed hexagonally into a tiling pattern bounded by any $N$-sided polygon(s) using the Filterbanking BeamFormer User Supplied Equipment (FBFUSE), an on-site, dedicated, high-performance cluster \citep{Barr2018}.
To do this, FBFUSE receives the channelised voltages from the correlator/beamformer's (CBF) F-engines \citep{VanDerByl2022} and executes multiple beamforming operations in real-time. The raw coherent and incoherent data from FBFUSE are written in the \texttt{SIGPROC}\footnote{\url{http://sigproc.sourceforge.net/}} filterbank format to the Accelerated Pulsar Search User Supplied Equipment \citep[APSUSE,][]{Barr2018}. The APSUSE backend is a high performance cluster housing 120 GPUs across 60 processing nodes.

Using interferometric beamforming for targeted pulsar searches has a number of advantages over previous survey methods. The `small diameter, long baseline' design of MeerKAT provides a wide field of view (FoV) on the angular scale of SNR diameters, while also providing accurate localisation of pulsars without the need for a timing solution. 
By compartmentalising searches into many CBs, the trade-off between field of view, telescope time and sensitivity are much more navigable. Additionally, multiple targets in the field can be searched simultaneously. At 1284~MHz, the sensitivity drops to 50 per cent at around 30~arcminutes from the FoV centre \citep{Asad2021}, setting a limit on the size of a source, or distance between sources, that can be observed in one instant. Beyond providing a flexible geometry, CBs are less susceptible to RFI \citep{Chen2021}. Furthermore, they do not `see' the full signal sky temperature contribution from the source, which is particularly useful for the ring shape of a SNR shell. This provides a sensitivity boost compared single-dish searches as the noise is reduced by a factor of approximately {$\sqrt{N_{\text{d}}}$ when $N_{\text{d}}$ telescopes are used, compared to the incoherent beam \citep[e.g.][]{Thompson2017}.}

\subsection{Sensitivity}
\begin{figure}
    \centering
    \includegraphics[width=1.0\columnwidth]{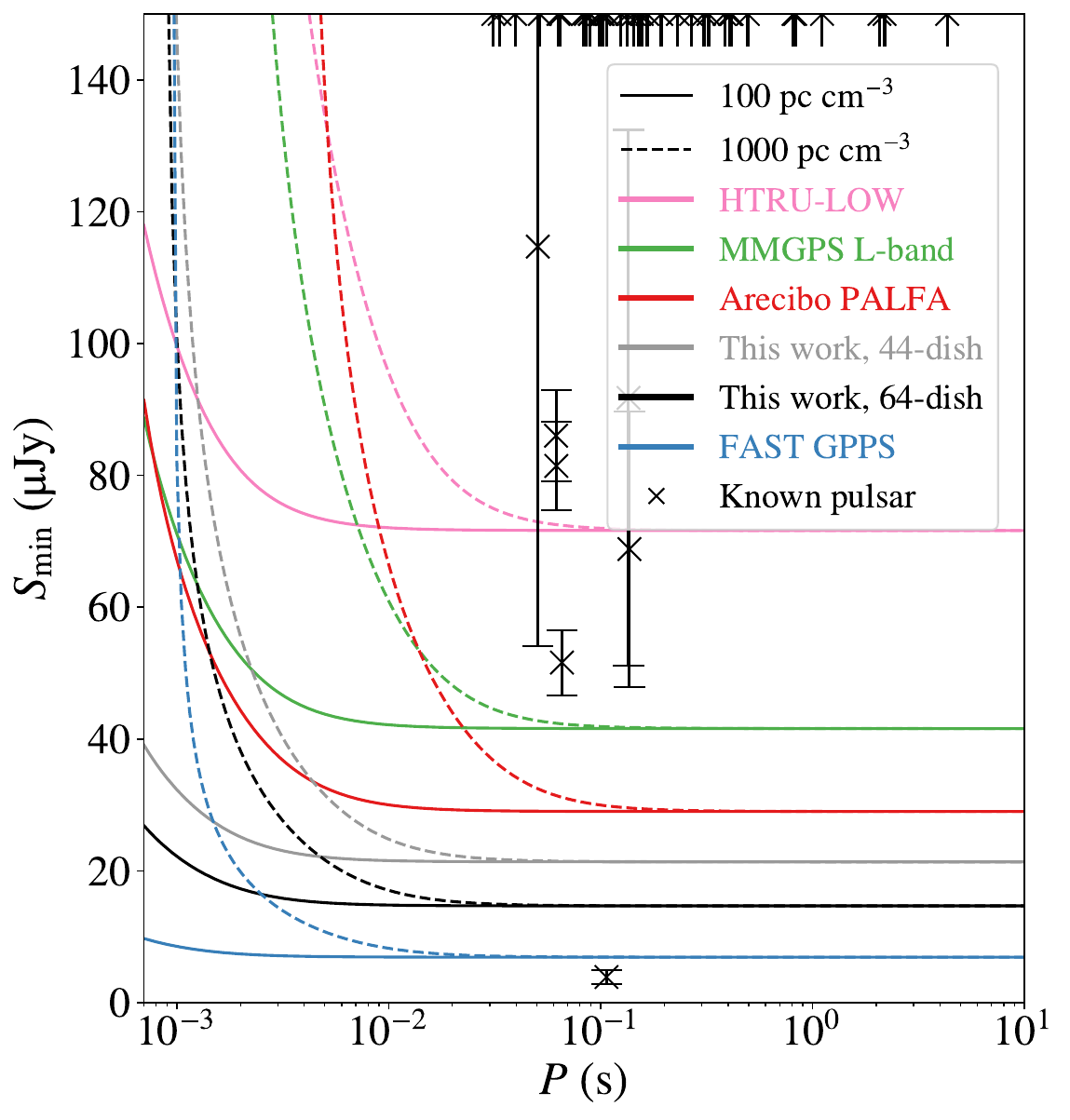}
    \caption{Lower limit for detectable pulse flux densities, \smin~as a function of pulse period for two representative DMs. $\epsilon$ is set to be unity and full bandwidth availability is assumed for all flux density lines. Two lines are shown for this work, each corresponding to the core 44-dish or full 64-dish array configuration. Also shown are the same sensitivity curves for the HTRU low-lat \citep{Keith2010}, PALFA survey with Arecibo \citep{Cordes2006}, MPIfR-MeerKAT Galactic Plane Survey at L-band \citep{Padmanabh2023} and the FAST Galactic Plane Pulsar Survey \citep{Han2021}. Flux densities of pulsars in SNRs and PWNe are plotted, though most have fluxes beyond the maximum shown and are therefore displayed as up-arrows. All curves and pulsar fluxes have been scaled to the central frequency of the MeerKAT L-band reciever, using a spectral index of $-$1.60$\pm$0.03. The error bars account for the uncertainty on the spectral index.}
    \label{fig:sens}
\end{figure}

The sensitivity of our searches is equivalent to the minimum flux density, \smin~we can detect for pulses of period, $P$ and width, $W$. It is described in terms of the system gain, $G$, temperature, $T_{\text{sys}}$, the observing time, $t_{\text{obs}}$ and bandwidth, $\Delta\nu$ by the modified radiometer equation \citep{Dewey1985} as
\begin{equation}
    S_{\text{min}} = \frac{\text{S/N} \, \beta \, T_{\text{sys}}}{\epsilon \, G \, \sqrt{N_{\text{p}} \, t_{\text{obs}} \, \Delta\nu }} \, \sqrt{\frac{W}{P-W}}.
    \label{eq:sens}
\end{equation}
We have appended an efficiency factor, $\epsilon$ of 0.7 as determined by \citet{Morello2020} to account for additional sensitivity loss from searching with a Fast Fourier Transform (FFT)-based search pipeline (see Section \ref{search}). $\beta$ is the factor accounting for the raw amplitude information lost during digitisation, which for 8-bit digitisation is $<$1 per cent \citep{Kouwenhoven2001}. For our calculations we assume this value to be 1. At L-band, the system temperature, $T_{\text{sys}}$ is a summation of the sky temperature, \tsky, the 18 K receiver temperature and the 4.5 K spillover contribution at 45 degrees elevation \citep{Bailes2020}. Assuming a minimum signal-to-noise, S/N of 9, 40 minute integration time and an upper bound on the duty cycle, $\delta\,=\,W/P$ of 10 per cent \citep{Posselt2023}, our observations achieve a flux density upper limit of 21~\uJy~or 31~\uJy~for the 64 or 44 dishes respectively. These calculations refer specifically to the centre of the coherent beam. The average dilution in sensitivity considering the pulsar could be located at any point in a coherent beam is around 35 per cent, though this varies depending on the baselines used and the position of the source in the sky. \autoref{fig:sens} displays the flux density sensitivity curve from \autoref{eq:sens} for this survey, and also for some non-targeted pulsar search surveys of the Galactic plane. For this survey, a \tsky~of 5.3~K is estimated by scaling the cold-sky temperature from \citet{Haslam1982} using a spectral index of $-$2.6 for diffuse radio synchrotron emission \citep{Reich1988,Zheng2017}. We scale the quoted \tsky~of the other survey publications to 1284~MHz in the same way. We set $\epsilon$~equal to 1 for all of the flux density curves in \autoref{fig:sens} to allow for a better comparison, as search pipeline efficiencies are sometimes not assigned for the non-MeerKAT surveys. 
The sensitivity degradation due to intra-channel dispersion smearing affects shorter pulse periods more strongly. The intrinsic width of the pulse, $W_{\text{i}}$ is broadened by sampling, scattering and DM smearing such that the measured width, $W^2\,=\,W_{\text{i}}^2\,+\,(\eta\,t_{\text{samp}})^2 + W_{\text{scatt}}^2\,+\,W_{\text{smear}}^2\,+\,W_{\text{DM}}^2\,$ \citep{Dewey1985}, where $\eta$ accounts for the finite pulse profile resolution due to the sampling time, $t_{\text{samp}}$, though for our searches we take it to be 1 \citep{Men2023}. For $W_{\text{scatt}}$, we use the expression derived by \citep{Lewandowski2015}, where $\text{log}\,W_{\text{scatt}} \text{(ms)}\,=\,-6.344\,+\,1.467\,\text{log DM}\,-\,0.509\,(\text{log DM})^{2}$. Our choice of 4096 channels ensures that at a DM below 1000\dm, the dominant factor limiting our sensitivity for $P > 10~$ms is \tsky. Also shown are the flux densities of pulsars associated with SNRs and/or PWNe for which this has been measured \citep{Manchester2005}, also scaled from 1400~MHz using the mean pulsar spectral index of $-$1.60$\pm$0.03 \citep{Jankowski2018}. The Galactic Plane Pulsar Snapshot (GPPS) survey \citep{Han2021} with the Five-hundred-meter Aperture Spherical Telescope \cite[FAST,][]{Li2016} achieves flux density limits of 7~\uJy, consistently more sensitive than our observations across all periods at which young pulsars are known to rotate. The sensitivity achieved by the PALFA survey with the Arecibo telescope \citep{Cordes2006} is fairly competitive with our searches for $P>$100~ms, but quickly deteriorates at higher DMs. Further to this, we expect many young pulsars in the Galactic plane to have fast periods in the range of 30-100~ms. We would expect to detect all 45 of the SNR-associated pulsars, with the exception of PSR J1907+0602, the pulsar with the lowest flux density in \autoref{fig:sens}. This pulsar was discovered following a 1.8~hr search with the Arecibo telescope \citep{Abdo2010a}. Consideration of such cases, where exceptionally deep observations uncovered a very faint pulsar, informed our decision to efficiently balance telescope time between the depth of each search and the number of targets.

\subsection{Sources targeted to date}\label{targets}
So far, we have observed 55 targets, of which 32 are G22 SNRs, 17 are candidate SNRs, 2 are isolated PWNe and 4 are unidentified TeV sources. Due to the extremely competitive sensitivity of the FAST GPPS, we avoid the footprint of that survey, therefore the targets of this survey are located at Galactic longitudes $<$30\degr. Information on each source and its observation are displayed in \autoref{tab:targets}. The sky temperature for each source was estimated from the GSM2016 model \citep{Zheng2017} using the \textsc{pygdsm}\footnote{\href{https://github.com/telegraphic/pygdsm}{https://github.com/telegraphic/pygdsm}} package \citep{Price2016} to obtain more accurate estimates of system temperatures for each source \citep{Price2021}. The sizes quoted are the extent of each source, not the explicit area searched with CBs. 

There is a trade-off in sensitivity for the more extended sources due to the CB limit. 
If a source is too large to be tiled with 780 beams, there are two options; i) change the CB overlap level or ii) increase the beam size by using only the innermost 44 dishes, which have a maximum baseline of 1 km. To calculate the number of beams required to fully tile a target, we use the multibeam simulation package \mosaic\footnote{\href{https://github.com/wchenastro/Mosaic}{https://github.com/wchenastro/Mosaic}{ by Weiwei Chen}}, described in \cite{Chen2021} to simulate the point spread function (PSF) of the CBs and their tiling positions. The PSF is time-dependent due to the movement of the source across the sky \citep{Chen2021}, so the PSF is approximated at the midpoint of the planned observation. We fix the radius at which the CBs overlap to be the 50 per cent power radius of the PSF. FBFUSE can only beamform using multiples of four dishes.

Among the first sources to be observed was SNR G28.7$-$0.4, which we wished to observe with the full 64-dish gain. Due to the reduced beam size of the 8~km baseline, we mitigated the prohibitively high data rate of 684 CBs by reducing the time resolution to 306~$\upmu$s. In all subsequent observations of more extended sources, however, we have optioned to use the core 44-dish baseline to provide a eight-fold increase in CB coverage, rather than halve the sampling rate which only increases the coverage by a factor of two.

All the sources in \autoref{tab:targets} can be covered by the half-power beam width of the incoherent beam, with the exception of G28.8+1.5. A handful of SNRs could not be fully tiled with coherent beams by just using the core baselines of MeerKAT. 
In these cases, which are indicated in \autoref{tab:specs}, we have targeted regions of interest. For G27.8+0.6 and G28.8+1.5, we placed coherent beams on two and seven X-ray point sources, respectively, that were identified by \citet{Misanovic2010} as potential PWN or NS candidates. The candidate remnant G350.8+5.0 \citep{Hurley-Walker2019b} contains the steep spectrum source GLEAM J170154$-$333857 which could be a pulsar. A circular region 21\arcmin~in diameter and centred on the GLEAM source has been searched. For the remnants G296.5+10.0, G347.3$-$0.5 and G353.6$-$0.7, we placed 480 CBs centred on the associated CCO, tiling a circular area 9.7\arcmin, 6.1\arcmin~and 9.1\arcmin~across, respectively, instead of the entire shell. G355.4+0.7 contains a possible associated \gray~source 4FGL~J1731.2$-$3235 \citep{Acero2016}, around which we tiled a region 5\arcmin~across with CBs. In future work we intend to produce maps of all beam tiling regions.

We have targeted a total of six CCOs (CXOU~J181852.0$-$150213 \citep{Reynolds2006}, 1E~1207.4$-$5209 \citep{Helfand1984, Zavlin2000}, 1WGA~J1713.4$-$3949 \citep{Slane1999}, XMMU~J172054.5$-$372652 \citep{Gaensler2008} and XMMU~J173203.3$-$344518 \citep{Halpern2010}) and three candidate CCOs (J174711.9$-$221741 \citep{Onic2019}, CXOU~J134124.22$-$634352.0 \citep{Hui2012} and CXOU~J171801.0$-$372617 \citep{Lazendic2005}). In these observations, the pointing position is chosen to be the best CCO position in the literature, not the geometric centre of the SNR. We do not neglect searching the rest of the SNR shell either, as there might be other sources overlapped by the shell that could be discovered.

\begin{landscape}
\begin{table}
\begin{threeparttable}
\caption{Sources searched so far in the survey and ordered by observation date. Some sources have alternative names which are provided at the bottom of the table. The type is defined as follows; `S' for shell-type G22 SNR, `C' for composite-type G22 SNR, `F' for filled centre G22 SNR, `P' for isolated pulsar wind nebula, `cand' for candidate SNR, `U' for unidentified TeV source. `?' denotes ambiguity in the literature. Source sizes are taken from the following; [1] \citet{Green2019} or \citet{Green2022} in the case of SNRs observed before or after 2023-09-01 respectively, [2] GLOSTAR \citep{Dokara2021}, [3] \citep{Brogan2006}, [4] MAGPIS \citep{Helfand2006},  [5] GLEAM \citep{Hurley-Walker2019b}, [6] the catalogue of TeV sources \texttt{tevcat} for H.E.S.S. sources, [7] \citep{Filipovic2022}. $S_{\text{min}}$ is calculated using \autoref{eq:sens}, where $\epsilon=0.7$, and specifically refers to the centre of a coherent beam. The upper limit on the radio luminosity of the pulses, $L_{\text{min}}\,=\,S_{\text{min}}D^{2}$, where the distance, $D$ is the most recent measurement. Previous flux density limits from targeted searches have been scaled to 1284~MHz using a spectral index of $-$1.6, and their original values are taken from the following; [8] \citet{Supan2015}, [9] \citet{Kaspi1996}, [10] \citet{Gelfand2007}. \\ $^{\dagger}$Target was observed simultaneously with one or more other sources, identifiable by their equivalent Observation Start times. \\ $^{a}$The associated object is a CCO or candidate CCO. \\ $^{b}$Due to the angular size of the source, only regions of interest were targeted with coherent beams. These are described in the text. \\ $^{c}$The associated object has a known period from X-ray timing.}
\label{tab:targets}

\begin{tabular}{lllllllllllll}
    \hline
    \textbf{Source} & \textbf{Type} & \textbf{Associated objects} & \textbf{Size} & \textbf{$t_{\text{obs}}$} & $t_{\text{samp}}$ & \textbf{Observation Start} & \textbf{$N_{\text{d}}$} & \textbf{$S_{\text{min}}$} & \textbf{Prev. $S_{\text{min}}$} & $D$ & & \textbf{$L_{\text{min}}$}  \\
     &  &  & (\arcmin) ref. & (s) & ($\upmu$s) & yyyy-mm-dd-hh:mm:ss &  & (\uJy) & (\uJy) ref. & (kpc) & & (mJy~kpc$^{2}$) \\
    \hline
G3.8+0.3 & S? &  & 18\hfill[1] & 2389 & 153 & 2023-04-23:07:10:19 & 44 & 38 & & 4.1 & \cite{Wang2020} & 0.6 \\
G005.161$-$0.321$^{\dagger}$ & cand &  & 11.2\hfill[2] & 1771 & 153 & 2022-10-13-09:58:05 & 60 & 33 & & & & \\
G005.378$-$0.280$^{\dagger}$ & cand &  & 4.4\hfill[2] & 1771 & 153 & 2022-10-13-09:58:05 & 60 & 34 & &  & &\\
G5.9+3.1$^{a}$ & S & J174711.9$-$221741 & 20\hfill[1] & 2387 & 153 & 2023-03-21-07:26:08 & 60 & 22 & &  & & \\
G11.0$-$0.0 & S & HESS J1809$-$193? & 11$\,\times$\,9\hfill[1] & 2383 & 153 & 2023-03-21-09:27:26 & 60 & 29 & & 2.4 & \cite{Shan2018} & 0.2 \\
G11.8$-$0.2$^{\dagger (i)}$ & S &  & 4\hfill[1] & 1787 & 153 & 2022-10-14-10:39:36 & 60 & 32 & & & & \\
G12.0$-$0.1$^{\dagger}$ & ? &  & 7\hfill[1] & 1787 & 153 & 2022-10-14-10:39:36 & 60 & 32 & & & & \\
G12.5+0.2 & C? &  & 6$\,\times$\,5\hfill[1] & 2390 & 153 & 2023-05-24-05:08:37 & 56 & 32 & & & &  \\
G12.7$-$0.0 & S &  & 6\hfill[1] & 1184 & 153 & 2022-10-14-11:30:31 & 60 & 42 & & & & \\
G013.500+0.074$^{\dagger}$ & cand &  & 3.8\hfill[2] & 1796 & 153 & 2022-10-13-11:18:46 & 60 & 34 & & & & \\
G13.5+0.2$^{\dagger}$ & S &  & 5$\,\times$\,4\hfill[1] & 1796 & 153 & 2022-10-13-11:18:46 & 60 & 33 & & 12.4 & \cite{Lee2020} & 5.0 \\
G15.4+0.1 & C? & HESS J1818$-$154 & 15$\,\times$\,14 \hfill[1] & 1084 & 153 & 2022-06-03-18:46:41 & 48 & 49 & 260 [8] & 8-10.5 & \cite{Su2017} & 3.1-5.4 \\
G15.51$-$0.15 & cand &  & 8$\,\times$\,9\hfill[3] & 1120 & 306 & 2022-02-07-03:16:40 & 60 & 40 & & & & \\
G15.9+0.2$^{a}$ & S & J181852.0$-$150213 & 7$\,\times$\,5\hfill[1] & 2389 & 153 & 2023-05-24-02:26:56 & 56 & 29 & & $>$7 & \cite{Tian2019} & $>$1.4 \\
G16.7+0.1 & C & PWN G16.73+0.08 & 4\hfill[1] & 1157 & 306 & 2022-02-07-03:36:28 & 60 & 39 & & 14 & & 7.6 \\
17.3361$-$0.1389$^{\dagger}$ & cand & G17.4$-$0.1? & 0.9\hfill[4]& 1500 & 153 & 2022-10-13-10:28:15 & 60 & 33 & & & & \\
G017.593+0.237$^{\dagger}$ & cand &  & 3\hfill[2] & 1500 & 153 & 2022-10-13-10:28:15 & 60 & 33 & & & & \\
G017.620+0.086$^{\dagger}$ & cand &  & 6\hfill[2] & 1500 & 153 & 2022-10-13-10:28:15 & 60 & 34 & & & & \\
18.1500$-$0.1722 & cand &  & 7\hfill[4] & 1125 & 306 & 2022-02-07-03:57:04 & 60 & 40 & & & & \\
G18.6$-$0.2 & S &  & 6\hfill[1] & 1188 & 153 & 2022-10-14-12:31:25 & 60 & 40 & & 4.4 & \cite{Ranasinghe2018} & 0.8 \\
G18.7583$-$0.0736 & cand &  & 0.8\hfill[4] & 1147 & 306 & 2022-02-07-04:17:00 & 60 & 41 & & & & \\
G20.0$-$0.2 & F &  & 10\hfill[1] & 1093 & 153 & 2022-06-03-18:06:41 & 48 & 49 & 131 [9] & 11.2 & \cite{Ranasinghe2018} & 6.2 \\
G021.492$-$0.010$^{\dagger}$ & cand &  & 7\hfill[2] & 1199 & 153 & 2022-10-13-10:58:25 & 60 & 37 & & & & \\
21.5569$-$0.1028$^{\dagger}$ & cand &  & 4\hfill[4] & 1199 & 153 & 2022-10-13-10:58:25 & 60 & 37 & & & & \\
G021.684+0.129$^{\dagger}$ & cand &  & 7.6\hfill[2] & 1199 & 153 & 2022-10-13-10:58:25 & 60 & 37 & & & & \\
G022.045$-$0.028 & cand & PWN G22.0+00.0? & 15.2\hfill[2] & 1087 & 153 & 2022-06-03-19:37:27 & 48 & 52 & & & & \\
G27.8+0.6$^{b}$ & S &  & 50$\,\times$\,30 \hfill[1] & 2386 & 153 & 2023-03-21-06:04:48 & 60 & 27 & 262 [9] & 4 & \cite{Wang2020} & 0.4 \\
G28.7$-$0.4$^{(ii)}$ & cand &  & 10\hfill[5] & 1148 & 306 & 2022-02-07-04:37:16 & 60 & 38 & & & & \\
G28.8+1.5$^{b}$ & S &  & 100?\hfill[1] & 2394 & 153 & 2023-05-24-03:07:19 & 56 & 25 & & 3.4 & \cite{Shan2018} & 0.3 \\
HESS J1844$-$030 & C? & G29.37+0.10 & 10.2\hfill[6] & 1189 & 153 & 2022-10-14-11:10:01 & 60 & 40 & & 6.5 & \cite{Petriella2019} & 1.7 \\
    \hline
\end{tabular}
\end{threeparttable}
\end{table}
\end{landscape}

\begin{landscape}
\begin{table}
\begin{threeparttable}
\contcaption{}
\label{tab:targetscont}
\begin{tabular}{lllllllllllll}
    \hline
    \textbf{Source} & \textbf{Type} & \textbf{Associated objects} & \textbf{Size} & \textbf{$t_{\text{obs}}$} & $t_{\text{samp}}$ & \textbf{Observation Start} & \textbf{$N_{\text{d}}$} & \textbf{$S_{\text{min}}$} & \textbf{Prev. $S_{\text{min}}$} & $D$ & & \textbf{$L_{\text{min}}$}  \\
     &  &  & (\arcmin) ref. & (s) & ($\upmu$s) & yyyy-mm-dd-hh:mm:ss &  & (\uJy) & (\uJy) ref. & (kpc) & &  (mJy~kpc$^{2}$) \\
    \hline
G031.256$-$0.041 & cand &  & 6.6\hfill[2] & 1189 & 153 & 2022-10-14-12:11:05 & 60 & 40 & & & & \\
HESS J1849$-$000$^{c}$ & P & PSR J1849$-$0001 & 1.7\hfill[6] & 1136 & 306 & 2022-02-07-04:57:36 & 60 & 34 & & 7 & \cite{Gotthelf2011} & 1.7 \\
ORC J0624$-$6948 & cand &  & 3.3\hfill[7] & 1145 & 153 & 2022-06-03-16:31:57 & 48 & 32 & & 50 & \cite{Filipovic2022} & 80 \\
G289.7$-$0.3 & S &  & 18$\,\times$\,14\hfill[1] & 1668 & 153 & 2022-06-03-19:48:18 & 48 & 31 & & & & \\
G296.5+10.0$^{abc}$ & S & PSR J1210$-$5226 & 90$\,\times$\,65\hfill[1] & 2379 & 153 & 2023-05-24-00:23:00 & 56 & 21 & 124 [9] & 1.4 & \cite{Eppens2024} & $<$0.1 \\
G298.6$-$0.0 & C? &  & 12$\,\times$\,9\hfill[1] & 1781 & 153 & 2022-06-03-16:53:37 & 48 & 31 & & 10.1 & \cite{Yeung2023} & 3.2 \\
G308.4$-$1.4$^{a}$ & S & J134124.22$-$634352.1 & 12$\,\times$\,6\hfill[1] & 2380 & 153 & 2023-06-26-23:27:57 & 60 & 22 & & 3.1 & \cite{Shan2019} & 0.2 \\
HESS J1427$-$608 & U &  & 3.3\hfill[6] & 2389 & 153 & 2023-05-24-01:44:29 & 56 & 25 & & & & \\
HESS J1507$-$622 & U &  & 9\hfill[6] & 2386 & 153 & 2023-04-23:04:29:05 & 44 & 28 & &  & \\
G318.9+0.4 & C &  & 30$\,\times$\,14\hfill[1] & 2378 & 153 & 2023-04-23:03:49:03 & 44 & 31 & & 3.5 & \cite{Wang2020} & 0.4 \\
G323.5+0.1 & S &  & 13\hfill[1] & 1092 & 153 & 2022-06-03-17:25:14 & 48 & 41 & & & & \\
G328.4+0.2$^{(iii)}$ & F? &  & 5\hfill[1] & 2394 & 153 & 2023-03-21-06:45:32 & 60 & 29 & 115 [10] & $>$17.4 & \cite{Gaensler2000} & $>$8.6 \\
G330.2+1.0$^{a}$ & S & J160103.1$-$513353 & 11\hfill[1] & 2388 & 153 & 2023-03-21-08:06:40 & 60 & 25 & & $>$4.9 & \cite{McClure-Griffiths2001} &$>$0.6 \\
G332.0+0.2 & S &  & 12\hfill[1] & 2395 & 153 & 2023-04-23:05:09:22 & 44 & 41 & 262 [9] & & & \\
HESS J1641$-$463 & U & G338.5+0.1 & 0\hfill[6] & 2377 & 153 & 2023-03-21-08:46:57 & 60 & 31 & & $>$11 & \cite{Kothes2007} & $>$3.7 \\
HESS J1708$-$410 & U &  & 4.8$\,\times$\,3.6 \hfill[6] & 2389 & 153 & 2023-05-24-03:48:01 & 56 & 29 & & & & \\
G347.3$-$0.5$^{ab}$ & S? & 1WGA J1713.4$-$3949 & 65$\,\times$\,55\hfill[1] & 2390 & 153 & 2023-05-24-01:03:48 & 56 & 28 & & 1.1 & \cite{Leike2021} & $<$0.1 \\
G349.7+0.2$^{a}$ & S & J171801.0$-$372618 & 2.5$\,\times$\,2\hfill[1] & 2396 & 153 & 2023-06-27-00:08:27 & 60 & 28 & & 11.5 & \cite{Tian2014} & 3.7 \\
G350.1$-$0.3$^{a}$ & S & J172054.5$-$372653 & 4\hfill[1] & 2391 & 153 & 2023-06-27-00:48:38 & 60 & 27 & & 4.5 & \cite{Borkowski2020} & 0.5 \\
G350.8+5.0$^{b}$ & cand &  & 72$\,\times$\,52\hfill[5] & 2390 & 153 & 2023-04-23:05:49:45 & 44 & 28 & & & & \\
G351.9$-$0.9 & S &  & 12$\,\times$\,9\hfill[1] & 2393 & 153 & 2023-04-23:06:30:03 & 44 & 35 & & & & \\
G353.6$-$0.7$^{a}$ & S & J173203.3$-$344519 & 30\hfill[1] & 2388 & 153 & 2023-06-27-01:28:52 & 60 & 25& & $<$3.5 & \cite{Wang2020} & $<$0.3 \\
G355.4+0.7$^{b}$ & S & 4FGL J1731.2$-$3235 & 25\hfill[1] & 2389 & 153 & 2023-05-24-04:28:16 & 56 & 30 & & 4.2 & \cite{Wang2020} & 0.5 \\
G357.7$-$0.1$^{(iv)}$ & S &  & 8$\,\times$\,3\hfill[1] & 2391 & 153 & 2023-06-27-02:09:05 & 60 & 31 & 262 [9] & 12 & \cite{Brogan2003} & 4.4 \\
    \hline
\end{tabular}
    \begin{tablenotes}
        \small
        \item $^{(i)}$11.8903$-$0.2250
        \item $^{(ii)}$G22.78$-$0.44
        \item $^{(iii)}$MSH 15$-$57
        \item $^{(iv)}$MSH 17$-$39, the Tornado
    \end{tablenotes}
\end{threeparttable}
\end{table}
\end{landscape}

\subsection{Timing observations}\label{Timing}
We continue to observe newly discovered pulsars in order to accurately measure their changing rotation rate. If the pulsar has been detected in more that one beam in the discovery observation, we can begin timing immediately by using \seekat\footnote{\href{https://github.com/BezuidenhoutMC/SeeKAT}{https://github.com/BezuidenhoutMC/SeeKAT}{ by Mechiel Bezuidenhout}} \citep{Bezuidenhout2023}, a multibeam localisation tool that uses the S/N and a PSF simulated by \mosaic, to obtain a position with arcsecond accuracy. If the pulsar is seen only seen in one beam, we attempt a multibeam detection using the first timing observation. The timing campaign is pseudo-logarithmic, consisting of nine epochs where the first two observations are spaced about half a day apart. This cadence is essential for keeping track of the rotation of the pulsar, as young pulsars tend to have a high rate of spin-down. A minimum of 58 of the 64 telescopes are used with FBFUSE and the APSUSE backend, to which the data are written for offline analysis.

We also make use of the UHF-band receivers for timing. The negative spectral index of pulsar radio emission may result in a higher signal-to-noise than at L-band for an equivalent integration time. The extended frequency range allows better opportunity for investigating frequency-dependent profile changes, which provide insights into the emission mechanism \citep[e.g.][]{Xu2021}. However, the effects of pulse broadening due to scattering and dispersive smearing are more pronounced in this band. To mitigate these effects, UHF observations use the Pulsar Timing User Supplied Equipment \citep[PTUSE,][]{Bailes2020} backend in search mode. PTUSE is a dedicated pulsar timing backend, providing much finer sampling in time and recording full Stokes polarisation information. Data are written in the \texttt{PSRFITS} format \citep{Hotan2004}, sampled every 7.5~$\upmu$s, recorded across 4096 channels and coherently dedispersed at the DM of the source. We did not use PTUSE for the L-band observations, because this could introduce a systematic offset in L-band pulse arrival times between the two backends.

The MeerKAT timing observations are supplemented with observations from Murriyang. These data are recorded with the Ultra Wide L-band (UWL) receiver covering 704-4032 MHz \citep{Hobbs2020}, and are stored in \texttt{PSRFITS} format. We use the search mode, channelising the data into 3328 channels across 26 subbands, and sampling every 64~$\upmu$s.

\section{Data Reduction}\label{search}
\subsubsection*{Searching}
To search for periodic signals, we use \textsc{peasoup}\footnote{\href{https://github.com/ewanbarr/peasoup}{https://github.com/ewanbarr/peasoup}{ by Ewan Barr}} \citep{Barr2020, Morello2019}, a GPU-based C++/CUDA library for time-domain linear acceleration searching for pulsations with frequencies between 0.1-1100~Hz. While we do not expect to find pulsar binaries in SNRs and wind nebulae, we elected to search accelerations between $\pm$5~m~s${^{-2}}$, assuming constant acceleration. A non-constant acceleration could result in a more accelerated pulsar being smeared below detection threshold if the entire integration time is searched at once, especially for tight-orbit binaries. We do not attempt to segment our searches as our integration times of 20-40~minutes would not risk much loss in sensitivity to mildly accelerated systems. Dedispersion is performed incoherently per DM trial using the \textsc{dedisp} library \citep{Levin2012, Barsdell2012}. As many sources in the survey lie on the Galactic plane, it is necessary to search up to a high DM. The contributions to pulse smearing due to DM step size and sampling time become dominated by the intra-channel smearing at higher DMs, so one can make use of a dedispersion plan that downsamples in DM and time to reduce the number of trials. We therefore generated a plan using \textsc{presto}/\texttt{DDPlan.py} \citep{Ransom2002, Ransom2011}, consisting of 8784 trials up to 3000\dm. Before being searched with \textsc{peasoup}, the data are cleaned using the \filtool~RFI cleaning algorithm from \pulsarx\footnote{\href{https://github.com/ypmen/PulsarX}{https://github.com/ypmen/PulsarX}{ by Yunpeng Men}} \citep{Men2023} which eliminates narrowband RFI using statistical thresholds and broadband RFI using a zero-DM matched filter \citep{Men2019}. It also replaces intra-channel samples of interference with noise. A separate channel mask is also applied, which masks 64 MHz of the band in order to mitigate common sources of narrowband RFI. After cleaning, the data are delivered to \textsc{peasoup} where they are dedispersed. For each DM trial, the cleaned and dedispersed time-series is de-reddened to remove low-frequency RFI before being re-sampled to account for the acceleration trial value. After this, the corrected series is Fourier transformed and harmonics are summed up to the 8$^{\text{th}}$ harmonic. To facilitate efficient use of the Fast Fourier Transform operation, the data are zero-padded from the final sample up to the next instance of 2$^{N}$. If the amplitude at a sampled frequency exceeds the S/N threshold, it is checked against a list of previously identified RFI frequencies before being written as a candidate.

The candidates, which normally number 10$^{3}$ per beam, are stored in XML format files and require sifting. We use \texttt{candidate\_filter.py}\footnote{\href{https://github.com/prajwalvp/candidate\_filter}{https://github.com/prajwalvp/candidate\_filter}{ by Lars Künkel}}, which performs further RFI mitigation and multibeam spatial clustering (a detailed description is found in \citealt{VoragantiPadmanabh2021}). This reduces the number of candidates by a factor of approximately 100.
The filtered candidates are folded with \texttt{psrfold\_fil}, another application of the \pulsarx~suite. \texttt{psrfold\_fil} optimises the period, period derivative and DM. The time-series to be folded is dedispersed using an algorithm that avoids redundant DMs, and is thus more efficient than brute force methods \citep{Men2023}. Optimised candidates are classified using two Pulsar Image-based Classification System (PICS) Machine Learning classifiers \citep{Zhu2014} that are trained on PALFA or TRAPUM data sets respectively. Candidates with a pulsar likelihood below a lenient score of 10 per cent are rejected. The final set of candidate diagnostic information and metadata are inspected by eye in the specialised candidate viewer CandyJar\footnote{\href{https://github.com/vivekvenkris/CandyJar}{https://github.com/vivekvenkris/CandyJar}{ by Vivek Venkatraman Krishnan}}. Candidates are designated either Tier 1 (high probability of novelty), Tier 2 (potentially novel but uncertain), RFI, noise or as a known pulsar. Upon identification of a Tier 1 candidate, we check that the discovery period is not a harmonic of the true period by refolding the filterbank using \texttt{psrfold\_fil} at integer multiples of the period; 2$P$ and 3$P$, and unit fractions; $P$/2 and $P$/3.

We investigate all the Tier 2 candidates by directly refolding the coherent or incoherent beam data. To do this, the data are cleaned with \texttt{filtool} then folded at the candidate's topocentric period with \dspsr\footnote{\url{https://dspsr.sourceforge.net/}}. A second round of cleaning is done using \clfd\footnote{\href{https://github.com/v-morello/clfd}{https://github.com/v-morello/clfd}{ by Vincent Morello}}. Finally, we use \pdmp~to optimise the S/N of the signal by searching over period, DM and acceleration. Both \dspsr~and \pdmp~are tools from the pulsar data analysis package \psrchive\footnote{\url{https://psrchive.sourceforge.net/index.shtml}}~\citep{Hotan2004,VanStraten2011}. The priors used by \pdmp~are informed by the uncertainty on the period, period derivative and acceleration of the candidate. The diagnostics are then inspected to see if the candidate is real. A higher S/N can potentially be extracted from the refolded data compared to the search due to the different optimisation methods; \pdmp~selects the highest S/N across every element in the parameter grid, whereas \texttt{psrfold\_fil} instead maximises the $\chi^{2}$ of the profile against the noise before calculating the S/N of the best profile. Any latent RFI that was not cleaned by \texttt{filtool} or \clfd~is flagged interactively using the \psrchive/\texttt{psrzap} tool to flag, and \psrchive/\texttt{psrsh} to mask the affected channels or subintegrations, which can further boost the S/N of the refolded data. 

Two sources in the survey are associated with neutron stars of a known rotational period but remain without a radio detection. The data of the coherent beam at the position of the neutron star is folded directly using an ephemeris from the known rotation parameters, and use both DM values predicted by the \textsc{ne2001} \citep{Cordes2004} and \textsc{ymw16} \citep{Yao2016} electron density models. To obtain the DM values, the \textsc{pygedm}\footnote{\href{https://github.com/FRBs/pygedm}{https://github.com/FRBs/pygedm}} \citep{Price2021b} package was used. The data are cleaned and folded as before, except this time \pdmp~searches over a DM range of $\pm$10~\dm. 

\subsubsection*{Timing}
Both the CB data from APSUSE and search mode PTUSE data are folded and cleaned by way of the same method as just described, but without searching in period or DM space, as these are already well constrained. Instead, the \psrchive/\texttt{pam} command is used to dedisperse and then combine the folded data in frequency and time to obtain the summed pulse profile. To generate the topocentric time-of-arrival (TOA) of the average pulse, the profile is cross-correlated with a noise-less profile template using the \psrchive/\texttt{pat} command. Noise-less templates are generated from the discovery observation using \psrchive/\texttt{paas}, in which we interactively fit one or more scaled von~Mises functions to return a high S/N mean profile. Separate templates are generated for each receiver band. For the template, we use either the profile of the discovery observation, or where possible, a high S/N summation of multiple profiles that have been aligned using a phase-connected timing solution.
With \texttt{TEMPO2}\footnote{\href{https://bitbucket.org/psrsoft/tempo2/src/master/}{https://bitbucket.org/psrsoft/tempo2/src/master/}} \citep{Hobbs2006}, the TOAs are corrected by referring them to the barycentre of the Solar System and fitted with a timing model. For UWL data from Murriyang, the same process for MeerKAT data is followed for cleaning folded data and generating TOAs. Any systematic offset in TOAs due to differing backends is corrected by fitting for a temporal jump.

\section{Results}\label{results}
Of the candidates inspected from the searches of all the sources in \autoref{tab:targets}, two Tier 1 candidates were identified and subsequently confirmed as new pulsars.

\subsection{PSR J1831\texorpdfstring{$-$}{-}0941}\label{sec:psrA}
\begin{figure}
    \centering
    \includegraphics[width=1.0\columnwidth]{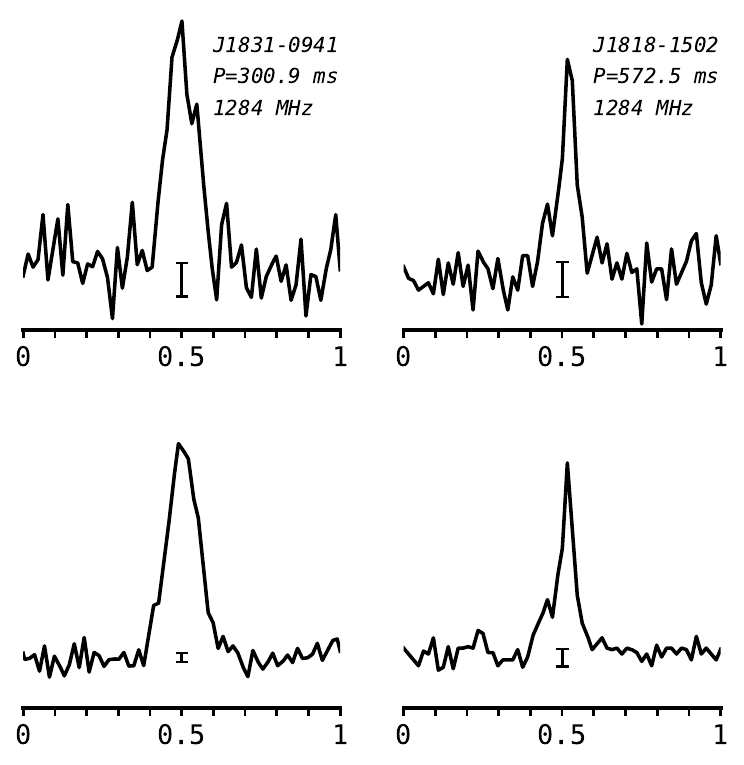}
    \caption{The integrated pulse profiles at L-band for the \psrA~discovery observation (upper left) and for combined timing observations (lower left) and equivalently for \psrB~(upper right and lower right, respectively). The data are folded onto 64 flux bins using the best ephemerides from timing, and are centred using the \psrchive/\texttt{pdv} centring function. The effective time resolution is greater than the sampling time due to DM smearing within frequency channels. This difference is less than one bin width for both pulsars but is nonetheless accounted for. The standard deviation of the noise is shown as an error bar at the bottom of each pulse. We see evidence for a precursor component for \psrB~at a phase of approximately 0.45.}
    \label{fig:profs}
\end{figure}

\begin{figure}
    \centering
    \includegraphics[width=1.0\columnwidth]{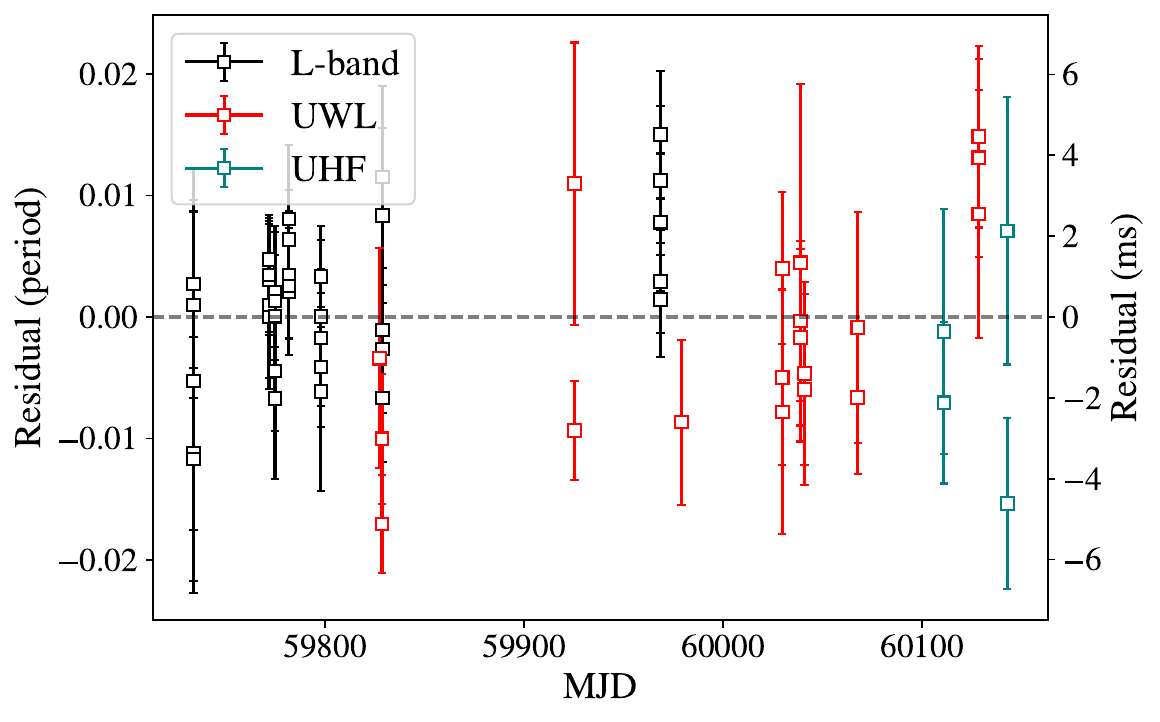}
    \includegraphics[width=1.0\columnwidth]{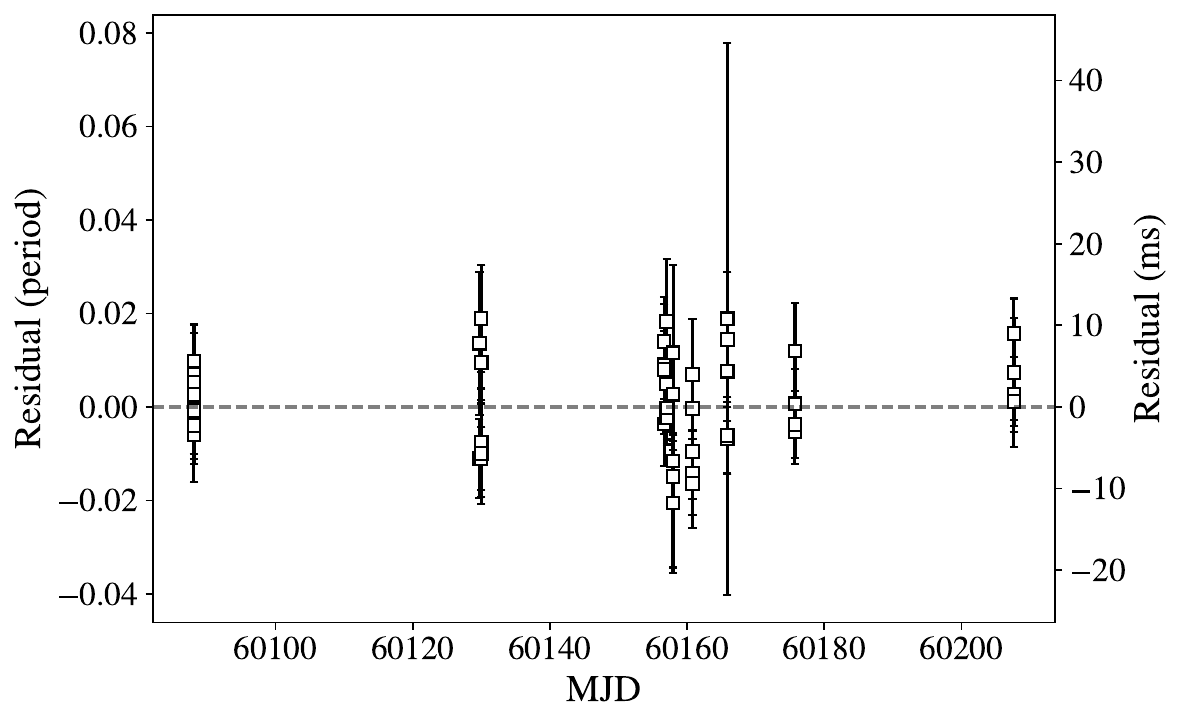}
    \caption{Residuals of the arrival times from the timing models for \psrA~(upper panel) and \psrB~(lower panel). The TOAs were fitted in \texttt{TEMPO2} resulting in the timing solutions in \autoref{tab:solns}. The folded data is split into multiple subintegrations in time. There are 58 TOAs shown for \psrA~and 49 for \psrB.}
    \label{fig:residuals}
\end{figure}

\psrA~was discovered in a coherent beam very close to the centre of G22.045$-$0.028, a candidate SNR identified in the GLOSTAR survey \citep{Dokara2021}, which itself encircles the candidate PWN G22.0+0.0 \citep[][hereafter YSB16]{Ueno2006, Yamauchi2016}. The pulsar appears to be isolated, and was detected at a topocentric rotation period of 300.85~ms, a dispersion measure of 377~\dm, and with a S/N of 16.8. The pulse profile generated from folding the data from this observation is shown in the upper left panel of \autoref{fig:profs}. \psrA~was detected in an additional 3 beams adjacent to the discovery beam. The S/N of the four profiles, each with 128 flux bins, was measured using the \pdmp~option of \psrchive/\texttt{psrstat} and using \seekat~we obtained a 1-sigma positional uncertainty of about 10\arcsec, or 1/6$^{\text{th}}$ the width of the CB.
\begin{table*}
\begin{center}
\caption{The properties of both new pulsars, including the measured and derived quantities from their coherent timing solutions after fitting TOAs in \texttt{TEMPO2}. The numbers in parentheses are the 1-sigma uncertainty on the final digit. The inferred distances are derived from the \textsc{ne2001} and \textsc{ymw16} electron density models. The upper limit on the period derivative for \psrB~is the 1-sigma uncertainty on the value returned when fitting for $\dot{P}$ in \texttt{TEMPO2}. The position for \psrB~has not been fitted, instead it is held at the \seekat~position.}\label{tab:solns}
\begin{tabular}{lll}
\hline \hline
Measured quantities & \psrA & \psrB \\ 
\hline
Right Ascension  \dotfill                 & 18$^\text{h}$31$^\text{m}$25\fs12(2) & 18$^\text{h}$18$^\text{m}$54\fs31$^{+0.16}_{-0.49}$ \\ 
Declination \dotfill & $-$09\degr41\arcmin53\farcs2(2)   & $-$15\degr02\arcmin04\farcs5$^{+4.7}_{-5.9}$ \\ 
Galactic co-ordinates, ($l$\degr, $b$\degr) \dotfill     & 22.030, $-$0.016 & 15.88, 0.19 \\
Period, $P$ (s) \dotfill & 0.3008716534(1)               & 0.5725490902(6) \\
Period derivative $\dot{P}$ (s s$^{-1}$) \dotfill        & $8.5612(6)\times10^{-14}$ & < $1.2\times10^{-16}$ \\
Duty cycle, $\delta$ (\%) \dotfill                       & 8.6(2) & 4.2(4) \\ 
Dispersion measure, DM (pc cm$^{-3}$) \dotfill           & 370.1(3) & 435(2) \\
Flux density, $S_{\mathrm{1284}}$ (\uJy) \dotfill        & 119(12) & 33(3) \\

\hline
Inferred quantities & & \\ 
\hline
Distance, $D$ (\textsc{ne2001}) (kpc) \dotfill & 4.9 & 5.4 \\
Distance, $D$ (\textsc{ymw16}) (kpc) \dotfill & 4.2 & 4.2 \\
Characteristic age, $\tau_c$ (kyr) \dotfill & 56 & > 78,000 \\
Surface dipole magnetic field strength, $B$ (G) \dotfill & $5.1\times10^{12}$ & < $3\times10^{11}$ \\
Spin-down luminosity, $-\dot{E}$~(erg~$\text{s}^{-1}$) \dotfill & $1.2\times10^{35}$ & < $8\times10^{30}$ \\
\hline \hline
\end{tabular}
\end{center}
\end{table*}
In total we have TOAs from 26 observations spanning 408 days. The residuals of these arrival times against the model of timing parameters in \autoref{tab:solns}, are shown in the upper panel of \autoref{fig:residuals}. The residuals have a root mean square (rms) of 2.2~ms. We measure a period derivative of 8.56~$\times$~10$^{-14}$~s~s$^{-1}$, from which we derive a characteristic age of 56 kyr and a spin-down luminosity of 1.2~$\times$~10$^{35}$erg~s$^{-1}$. This places \psrA~within the population of young pulsars in SNRs.
\subsection{PSR J1818\texorpdfstring{$-$}{-}1502}
\begin{figure*}
    \centering
    \includegraphics[width=0.45\textwidth]{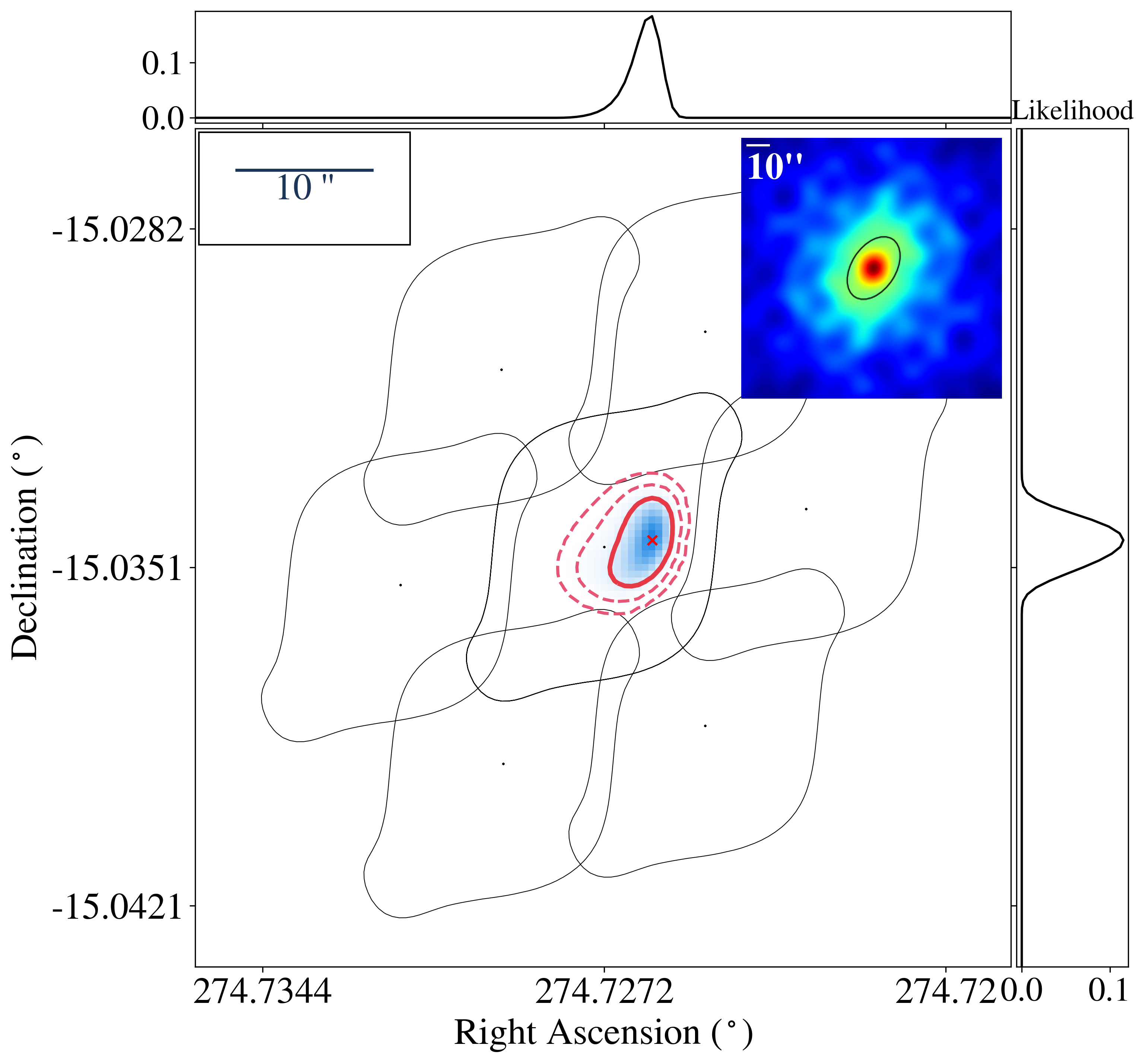}
    \includegraphics[width=0.5\textwidth]{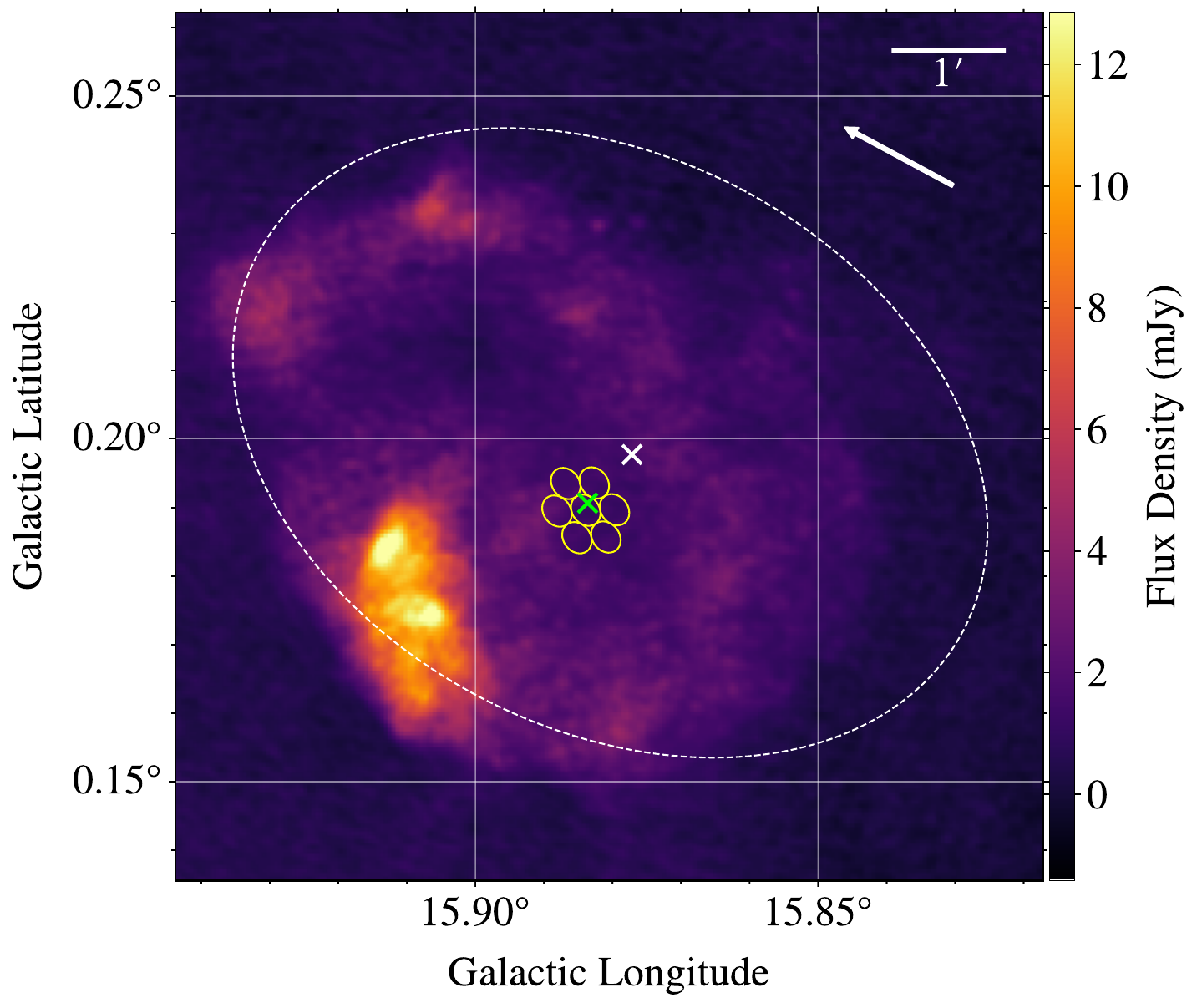}
    \caption{\textit{Left:} Close up of the localisation from \seekat. The red contours are the two (bold), three and four-sigma levels of the position likelihood. The likelihood function is plotted on the opposite sides of the position axes. The 50 per cent power level contour of the PSF simulated by \mosaic~is plotted for each beam as solid black lines. The PSF is displayed in the upper right with an elliptical approximation of the 50 per cent power contour. \textit{Right:} Image of G15.9+0.2 from MAGPIS at 20cm with the VLA \citep{White2005}. The seven beams in which \psrB~was detected are shown in yellow, and the best position from \seekat~is marked with a green cross. The white dashed ellipse defines the shape of the remnant \citep[G22,][]{Dubner1996}. The white cross is the position of CXOU J181852.0$-$150213 from \citet{Mayer2021}. The white arrow points North.}
    \label{fig:J1818_loc}
\end{figure*}

\psrB~was discovered in a coherent beam with a S/N of 14.6 in a search of the shell-type remnant G15.9+0.2. It has a period of 572.5~ms and a DM of 435~\dm. The profile of \psrB~from this observation was produced in the same way as for \psrA, and is shown in the upper right panel of \autoref{fig:profs}. Using the same procedure as in Section \ref{sec:psrA}, \seekat~returned a 4-sigma position of 18$^\text{h}$18$^\text{m}$54\fs3$^{+0.2}_{-0.5}$, $-$15\degr02\arcmin04\farcs5$^{+4.7}_{-5.9}$. The probability distribution of the localisation is shown in the left-hand side of \autoref{fig:J1818_loc}. The pulsar's position within G15.9+0.2 is shown in the map on the right-hand side. The quoted uncertainty from \seekat~is more representative of the total position error than 3-sigma due to the low signal-to-noise of the detections \citep{Bezuidenhout2023}.

As of now we have 8 observations spanning 118 days, for which TOAs have been generated using a template made from the discovery observation. The timing residuals from a fit of the \seekat~position, and the period and period derivative in \autoref{tab:solns}, are shown in the lower panel of \autoref{fig:residuals}. They have a rms of 7.6~ms. We are not yet able to make a significant measurement of $\dot{P}$, and so present an upper limit equivalent to the 1-sigma uncertainty of 1.2$\,\times\,10^{-16}$~s~s$^{-1}$ on the fitted values. We are confident that the actual magnitude of the spin-down rate is much smaller than this value, as it would otherwise have been resolved with the cadence and baseline of these TOAs. The upper limit on $\dot{P}$ from the fit leads us to conclude that \psrB~is a very old canonical pulsar with a characteristic age greater than 78~Myr. We will continue timing \psrB~in order to determine the true $\dot{P}$ and constrain the age, spin-down luminosity and magnetic field strength. It is possible that in fitting for $\dot{P}$, we are absorbing an apparent change in period due to a position error. The maximum position error at 4-sigma is 7\arcsec. If the true position differs by this amount then the apparent change in rotation period, if the period were measured at the near and far side of the Earth's orbit, induces a period derivative of 8$\,\times\,10^{-17}$~s~s$^{-1}$, which is the same order of magnitude the uncertainty of the fit. Therefore a baseline of at least one year will be required to ensure the position and period change of the source are disentangled.
\subsection{Pulse profiles}
The widths of the profiles at 50 per cent of the peak intensity, W50 at 1284~MHz have been calculated using a high S/N profile of each pulsar by combining data from all MeerKAT observations where a coherent beam was placed directly on the best position of the pulsar. For \psrA~and \psrB~there are 6 and 7 such observations, respectively. The pulse profiles of these high signal-to-noise data are shown in \autoref{fig:profs} where the lower left panel the profile of \psrA~and the lower right panel shows that of \psrB. We obtained these profiles by folding the data from each timing observation with \dspsr, using ephemerides composed of the measured parameters in \autoref{tab:solns}. They were cleaned with \clfd~before being summed in frequency and time. \psrchive/\texttt{psradd} was used to combine the profiles into an average profile. To calculate the width of each profile, a noise-less template was generated by modelling the pulse component(s) of the profile with von~Mises functions in \psrchive/\texttt{paas}. To determine W50, the template had noise added to it using the noise statistics of the average profile's off-pulse region, before being fitted using the \texttt{fitvonMises} function in \texttt{PSRSALSA} \citep{Weltevrede2016} which considers the concentration parameter of the best combination of von~Mises fits. This process is repeated 1000 times. Using the concentration parameter of the \texttt{fitvonMises} function from all 1000 fits, we get a distribution of widths and obtain W50 and its uncertainty from the mean and standard deviation. The mean widths from each distribution are W50\,=\,(25.7\,$\pm$\,0.5)~ms for \psrA, and W50\,=\,(24.1$\,\pm\,$2.7)~ms for \psrB. These are converted to a duty cycle, $\delta$ and presented in \autoref{tab:solns}. The pulse width for \psrA~is larger than most pulsars of a similar period \citep{Posselt2021}.

The pulse shape of \psrA~appears single-peaked and approximately symmetric, whereas \psrB~appears to have a main pulse preceded by a weaker and slightly narrower precursor. The evidence for the reality of the precursor is two-fold. Two von~Mises functions are required to adequately model the profile of \psrB~in \psrchive/\texttt{paas}. Secondly, the distribution of W50 from the iterative profile fits is not Gaussian; aside from a peak at approximately 25~ms, there is another peak at about 21~ms, which likely corresponds to narrower von~Mises functions not capturing the precursor. This asymmetry inflates the uncertainty on $\delta$.

We investigated the effect of interstellar scattering on the profile of \psrA~at L-band. The same high signal-to-noise profile with which we measured the pulse width was used. The scattering analysis tool \textsc{scatfit}\footnote{\href{https://github.com/InesPM/scatfit}{https://github.com/InesPM/scatfit}{ by Fabian Jankowski}} \citep{Jankowski2023} was used to split the 856~MHz band into four subbands, each of 256 channels, and fit an exponentially modified Gaussian distribution to the flux. This function is a robust description of a pulse scattered by a thin scattering screen \citep{Cordes2001a, Oswald2021}. We find little evidence for pulse broadening due to scattering across the band. A scattering timescale of (12$\,\pm\,$4)~ms is measured between 856-1070~MHz. The dispersion smearing timescale dominates broadening at higher frequencies.

The flux densities of the new pulsars are calculated using \autoref{eq:sens} where $W\,=\,\delta P$, omitting $\epsilon$ and using $\Delta\nu_{\text{eff}}$ from \autoref{tab:specs}. The sky temperatures at both positions were determined using the GSM2016 model \citep{Zheng2017}. We find the flux density at 1284~MHz, $S_{1284}$ to be (119$\,\pm\,$12)~\uJy~for \psrA, and (33$\,\pm\,$3)~\uJy~for \psrB. The uncertainties on these values of 10 per cent are determined from the error on both the sky temperature and the signal-to-noise. Both are fainter than could reasonably be detected by the vast majority of previously targeted pulsar searches of SNRs. Using the DM-derived distances in \autoref{tab:solns}, \psrA~has a radio luminosity, $L_{1284}$ of 2.1-2.9~mJy~kpc$^{2}$, whereas \psrB~has a luminosity of just 0.6-0.8~mJy~kpc$^{2}$. The range is the difference between the \textsc{ne2001} and \textsc{ymw16} models. \psrB~has a pseudo-luminosity comparable to some of the more intrinsically faint pulsars that have been found by the FAST GPPS survey \citep{Han2021}.

\subsection{Non-detections}\label{nondetections}
Many of the observations produce a handful of Tier 2 candidates. So far, none of these have been found to be convincing signals from new pulsars after follow up analysis described in Section \ref{search}. We regularly detect known pulsars present in the FoV, mainly in the incoherent beam, but occasionally in clusters of coherent beams at the edge of the tiling arrangements. Indeed, analysis by \citet{Padmanabh2023} find that known pulsars are detected as expected in the MPIfR-MeerKAT Galactic Plane Survey at L-band, which also uses the \textsc{peasoup} pipeline. The upper limits on the targets we have searched are 21-52~\uJy~at 1284~MHz, and are a significant improvement on the depth of previous targeted searches or of non-targeted searches of the plane.

\section{Discussion}\label{discussion}
\begin{figure}
    \centering
    \includegraphics[width=1.0\columnwidth]{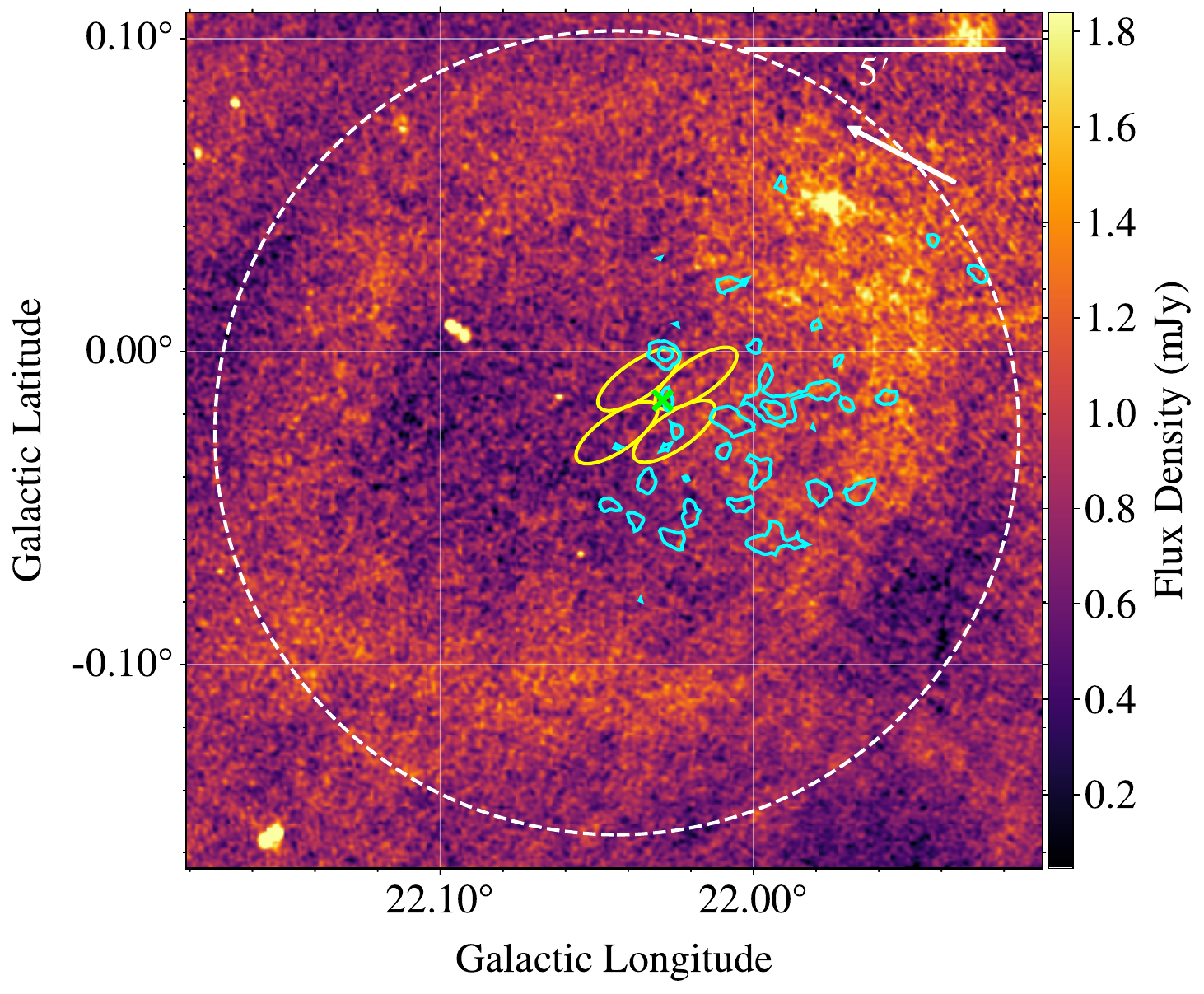}
    \caption{Image of the field towards G22.045$-$0.028 from MAGPIS at 20cm with the VLA \citep{White2005}. The blue contours are the X-ray emission seen by YSB16 (reprocessd Suzaku data). The yellow ellipses and green cross are the four discovery beams and best position of \psrA, respectively. The dashed circle is the shape of G22.045$-$0.028 \citep{Dokara2021}. The white arrow points North.}
    \label{fig:22.0+0.0_field}
\end{figure}
\subsection{Association between PSR~J1831\texorpdfstring{$-$}{-}0941, G22.0+0.0 and G22.045\texorpdfstring{$-$}{-}0.028}
G22.0+0.0 was first identified by \citet{Ueno2006} during a reanalysis of \textit{ASCA} Galactic Plane Survey \citep{Yamauchi2002} data to search for synchrotron emission from low latitude SNRs with a low radio surface brightness \citep{Sugizaki2001}. They identified an extended region of diffuse, hard X-ray emission about 10$'$ across. 
YSB16 re-detected the emission and measured the distance, for which they used Galactic HI column density and CO velocity measurements, to be 5.2$\pm$2.2 kpc. The uncertainty range overlaps with the DM-derived distances to \psrA~in Table \ref{tab:solns}. The discovery of G22.045$-$0.028 \citep{Dokara2021}, a 15.4\arcmin~across shell encompassing the X-ray position, provided more evidence for a plerionic scenario. The position of \psrA~relative to the radio shell and X-ray emission is shown in \autoref{fig:22.0+0.0_field}. We used the reprocessed\footnote{\textit{Suzaku} data is available at to browse from HEASARC at \href{https://heasarc.gsfc.nasa.gov/cgi-bin/W3Browse/w3browse.pl}{https://heasarc.gsfc.nasa.gov/cgi-bin/W3Browse/w3browse.pl}} \textit{Suzaku} data from the observation by YSB16 (Obs ID: 505025010). Using the \texttt{xselect}\footnote{\href{https://heasarc.gsfc.nasa.gov/docs/software/lheasoft/ftools/xselect/}{https://heasarc.gsfc.nasa.gov/docs/software/lheasoft/ftools/xselect/}} command from \textsc{ftools} \citep{HEAsoft2014}, the data from the three detectors were added and then smoothed using a 1-sigma Gaussian kernel.

YSB16 estimated a characteristic age of 10$^{4}$-10$^{5}$~yr for the putative pulsar, based on its clumpy morphology and also the relation between age and size in the X-ray band \citep{Bamba2010}, to which our characteristic age of 55.7~kyr agrees. The position, inferred distance and adolescence of \psrA~provides strong evidence that G22.0+0.0 and G22.045$-$0.028 form one composite-type SNR. 

At 5.2$\pm$2.2~kpc and 10\arcmin~across, the nebula is 15$\pm$6~pc in diameter, and the shell is 23$\pm$9~pc across. If we assume the age of the remnant is equivalent to $\tau_{\text{c}}$ from \autoref{tab:solns}, then we find G22.0+0.0 follows the observed relation between X-ray PWN size and age \citep{Bamba2010}, which predicts a size of 10-21~pc. Given an X-ray flux from YSB16 of 8.36~$\times$~10$^{-9}$~erg~s$^{-1}$~m$^{-2}$, we derive an X-ray conversion efficiency of 2.2\% at 5.2~kpc, which is higher than other pulsars of a similar $\dot{E}$ \citep{Possenti2002}, though contained within the spread of values for any given $\dot{E}$.
A high photon index of the X-ray spectra is expected for more efficient winds, as the spectrum of the radiation depends on the injected energy spectrum and wind magnetisation, which themselves are driven by the spin-down power of the pulsar \citep[e.g.][]{Torres2013}. The photon index of 1.7$\pm$0.3 measured by YSB16 is typical for PWNe and agrees with this trend \citep{Kargaltsev2008}.

As \autoref{fig:22.0+0.0_field} shows, \psrA~is located centrally within the shell whilst the nebula is concentrated in the South-Western quadrant. Given the estimated age of the system, it is probable that the nebula has already been disrupted by the reverse shock \citep{Gelfand2009}, which would have temporarily counteracted the expansion of the nebula and broken up the wind as it mixed with the reheated interior \citep{Blondin2001}. It is possible that this process is ongoing, which would perhaps explain the higher than average conversion efficiency as a consequence of a surge in pressure and magnetisation of the pulsar wind. The displacement of the PWN could be explained by a westwards gradient in the ISM density, making G22.0+0.0 a similar system to SNR G327.1$-$1.1 \citep{Temim2015}. The cluster of H\textsc{ii} regions to the West of the shell \citep{Dokara2021} could be related to the density asymmetry.

$\tau_{\text{c}}$ is not a fully reliable proxy for the true age, a point which is discussed this further ahead in this section. If, for now, we use $\tau_{\text{c}}$ , we calculate the transverse component of the kick velocity, $v_{\perp}$. The geometric centre of the G22.045$-$0.028 from \citet{Dokara2021} is assumed as the explosion site. The PWN is not an indicator of birth location, due to the effects of the reverse shock interaction and the movement of the NS. Assuming a single system, we use the distance to the PWN from YSB16, we find $v_{\perp}$~=~32~\kms, with a range of 18~\kms~to 45~\kms~given the $\pm$2.2~kpc~uncertainty. Depending on the magnitude of the radial velocity, the $v_{\text{kick}}$ vector is likely small compared to the young pulsar population \citep{Hobbs2005}. The observed change in period induced by $v_{\perp}$ across the celestial sphere, known as the Shklovskii effect \citep{Shklovskii1970}, would increase the observed period derivative by 7~$\times$~10$^{-21}$, thus is not significant compared to the uncertainty on the $\dot{P}$ we have obtained. We have investigated whether a bow-shock scenario is realistic. No radio emission, nor any GeV or TeV excess has been identified at position of the PWN. However, it is not unusual for bow-shock nebulae to only show up in one frequency regime due to the different radiative cooling timescales. Pulsars within shells are not expected to cause bow-shocks due to the low density of the medium after being swept up by the SN shock, nevertheless it is possible the pulsar may have escaped the far side of the remnant. A pulsar with $v_{\text{kick}}~\geq$~650~\kms~would be able to escape its shell before the remnant dissipates \citep{VanDerSwaluw2003}, and in doing so would form a bow-shock that emits copious radio synchrotron photons \citep{Shull1989}. We note both the lack of observational evidence for a bow-shock scenario, and also the small likelihood of a dominant radial component of $v_{\text{kick}}$ that would be required. Thus, we do not consider G22.0+0.0 to be a bow-shock nebula.
\citet{Dokara2021} were not able to identify any radio emission from the wind nebula itself. We can estimate the expected angular size of this emission to determine whether it is not luminous enough, or has been resolved out. \citet{Stappers1999} derived the angular size of the region shocked by the wind to be
\begin{equation}\label{eqn:pwn_size}
    \theta = \frac{1\farcs05}{D_{\text{kpc}}\,v_{1000}}\,\sqrt{\frac{f}{n_{0}}},
\end{equation}
where $D_{\text{kpc}}$ is the distance in kpc, $v_{1000}$~=~v$_{\text{kick}}$~$\times$~1000~\kms, $f$ is the energy conversion fraction from the spin-down energy to the wind and $n_{0}$ is the local number density of particles in cm$^{-3}$. If we assume a spherical wind in the face of reverse shock disruption, and also assuming $v_{\text{kick}}~\approx~v_{\perp}$, $n_{0}$~=~0.1 and $f$~=~1\%, then $\theta$~=~2\arcsec, much smaller than the 18\arcsec~resolution of the GLOSTAR observations. There is no mention of a point source, thus we can estimate an upper limit on the radio luminosity, using the methodology of \citet{Frail1997} and scaling the frequency using a Crab-like spectral index of $-$0.3. Given the upper value for the rms noise is 0.15~mJy at an effective central frequency of 5.8~GHz \citep{Dokara2021}, the radio luminosity, $L_{\text{R}}$ given by
\begin{equation}\label{eqn:pwn_luminosity}
    L_{\text{R}} = 9.1\,\times\,10^{28}\,D_{\text{kpc}}^{2}\,S_{5.8}
\end{equation}
is 3.7~$\times$~10$^{29}$~mJy~kpc$^{2}$, corresponding to a conversion efficiency, $L_{\text{R}}$/$\dot{E}$ of about 3~$\times$~10$^{-6}$. This is a higher efficiency upper limit than the peak of the distribution from non-detections in the study of pulsars with 31.88\,$\leq$\,log($\dot{E}$)\,$\leq$\,36.53 by \citet{Frail1997}, which is not unexpected given that the radio nebula luminosity appears to fall with age. It is possible that the nebula may be a TeV source, but falls below current sensitivity limits given the distance and 
age of the pulsar, though may be detectable by next generation TeV observatories such as the \v{C}erenkov Telescope Array \citep[e.g.][]{Mitchell2022}.

As mentioned, these calculations come with the caveat that the model-dependent distances are subject to large and unquantified uncertainties, and that the characteristic age is not a robust evaluation of the true age. $\tau_{\text{c}}$ is contingent on the assumption that the pulsar spin-down is purely driven by the magnetic dipole radiation, and thus the braking index, $n$ is equal to 3, where $\dot{\Omega}\,\propto\,\Omega^{n}$ \citep{Manchester1977}, where $\Omega$\,=\,1/$P$. It also assumes the birth period is much less than the present day period. A realistic birth period of 30~ms would only cause $\tau_{\text{c}}$ to change by one per cent for \psrA. A much larger uncertainty is imposed by the reality that $n<$\,3 \citep{Noutsos2013} and may not be constant. For $n$\,=\,2.5 \citep{Lyne2015}, the true age would be 30 per cent greater that $\tau_{\text{c}}$. Apart from seeking an age from which to derive $v_{\text{kick}}$, an accurate velocity could instead be obtained from measuring the proper motion. The contribution due to the natal kick would be only about 13~mas~year$^{-1}$, inducing an apparent $\dot{P}$ of 8~$\times$~10$^{-21}$~s~s$^{-1}$, corresponding to a TOA drift of less than 10$^{-4}$ turns per year.

It is reasonable to expect that high $\dot{E}$ pulsars might be detectable at \grays, with one reason being that the wider beams in the VHE band gives a greater sky coverage, perhaps a factor of 2 higher than radio beams, even for pulsars of moderately high $\dot{E}$ \citep{Ravi2010}. 

\psrA~lies within the localisation region of a \textit{Fermi} Large Area Telescope (LAT) \gray~source, 4FGL~J1830.8-0947 \citep{4FGL,4FGLDR4}, however this source has already been identified as being powered by the energetic PSR~J1831$-$0952 \citep{Laffon2015}. It is possible that \gray~emission from \psrA~is masked by this source, but unfortunately the validity period of the timing solution and the precision on the spin parameters are insufficient for applying the methods of \citet{Bruel2019} and \citet{Smith2019} to search for \gray~pulsations in the absence of a possibly associated point source. \gray~pulsations from \psrA~may yet be detected in the \textit{Fermi}-LAT data with a future, more precise timing ephemeris, and with a detailed study of the \gray~emission from this busy region of the Galactic plane. 
As a young pulsar, \psrA~is a member of a population of pulsars known to glitch, i.e. suddenly increase their spin frequency counter to their spin-down. Pulsars are more likely to glitch when young, possibly because the higher spin-down rate is conjunctive to the trigger mechanisms that might be responsible for disrupting the angular momentum distribution of the NS interior \citep[see e.g.][]{Antonopoulou2022a}. We see no evidence for a glitch of any magnitude in the timing residuals we have obtained so far. Using the power-law relation between $\tau_{\text{c}}$ and the glitch rate from \citet{Basu2022}, \psrA~could be expected to glitch at a rate between 0.2-0.4~yr$^{-1}$ using the 1-sigma uncertainty range. We would therefore expect to wait 4.5~yrs before we could expect a glitch to occur with a confidence of 68 per cent. Furthermore, one may also expect a handful of glitches within the 15-yr interval spanned by the \textit{Fermi}-LAT data, which would further complicate efforts to detect \gray~pulsations from.

\subsection{PSR J1818\texorpdfstring{$-$}{-}1502}
G15.9+0.2 is a shell-type SNR of small diameter first identified in the radio band \citep{Clark1973}. The detection of a radio pulsar at this position is interesting given that the SNR already has an associated neutron star. The central compact object CXOU J181852.0$-$150213, also shown in \autoref{fig:J1818_loc} is best candidate for the collapsed core of the progenitor star the assumed explosion site \citep{Mayer2021}. We see no evidence of radio pulsations at the position of the CCO to an upper limit of 29~\uJy. The distance to \psrB~from a DM of 435~\dm~is 5.4~kpc (\textsc{ne2001}) or 4.2~kpc (\textsc{ymw16}) though these are subject to large uncertainties. The distance to the shell is 7-16~kpc from HI absorption \citep{Tian2019}, suggesting that \psrB~is an unrelated pulsar in the foreground.

The CCO is offset by about 0.6\arcmin~from the geometric centre of the SNR at 18$^\text{h}$18$^\text{m}$54\fs2, $-$15\degr01\arcmin55\arcsec. However, the pulsar is much closer to this position at only 0.2\arcmin~away, and only 0.5\arcmin~from the CCO. The height of \psrB~above the Galactic plane would be 23~pc at the lowest estimate of the distance to G15.9+0.2 of 7~kpc. If we take the scale height for pulsars with a characteristic age of $>$78~Myr to be 400-600~pc \citep{Mdzinarishvili2004}, approximately 95 per cent of pulsars as old as \psrB~would lie $>$23~pc from the Galactic plane. However, the Galactic plane is densely populated with known pulsars, and becoming increasingly more so, thus it is not completely improbable that we should discover an old pulsar in this part of the sky. We therefore note that despite the offset between the CCO and \psrB~being unusually small, this does not by itself suggest a evolutionary connection between the two. 
\subsection{Upper limits on sources with known periods}
Two of our targets are young neutron stars with known rotation periods measured from X-ray timing. One is the 38.5~ms radio-quiet pulsar PSR J1849$-$0001 \citep{Gotthelf2011}. This pulsar emits non-thermal X-rays and powers the wind nebula IGR~J18490$-$0000/HESS~J1849$-$000. The nebula has been detected at both X-ray \citep{Calas2018} and VHE \gray~\citep{Terrier2008, Abdalla2018a} wavelengths. This is an archetypal PWN system but not easily studied due to its faintness at a distance of about 7~kpc \citep{Gotthelf2011}. We speculated that the distance and proximity to the Galactic plane may have inhibited the detection of radio pulsations of previous surveys, for which our search would be more resilient due to the sensitivity and wide bandwidth extending above 1700~MHz where scattering effects are weaker. However, to a flux limit of $S_{\text{min}}~=$~34~\uJy, we do not see any pulsations from PSR J1849$-$0001. This could be due to a differing geometry between the X-ray and radio beams, such that the radio beam is directed too far away from the line of sight to Earth to be detectable. Alternatively, the radio beam may be incident but the emission too faint due to a insufficient radio conversion efficiency, $\eta_{\text{R}}$. We can estimate an upper limit on $\eta_{\text{R}}$ using our flux limit and the relation between luminosity and flux density from \citet{Lorimer2005}, where the radio luminosity, $L_{\text{R}}\,=\,2.7\,\times\,10^{27}\,(\frac{S_{1400}}{\text{mJy}})^{2}\,(\frac{d}{\text{kpc}})^{2}$~erg~s$^{-1}$. Using the estimated distance of 7~kpc, it follows that the upper limit for $\eta_{\text{R}}\,=\,\frac{L_{\text{R}}}{\dot{E}}\,=$~(3$\pm$1)\,$\times$\,10$^{-11}$, where $S_{\text{min}}$ has been scaled from 1284~MHz to 1400~MHz using a spectral index of $-$1.6. A distance uncertainty of 1~kpc was assumed to account for the depth of the Scutum-Centaurus Arm in which the system is thought to located. At 7~kpc, the X-ray conversion efficiency, $\eta_{\text{X}}$ is $\approx$0.003 for flux integrated between 2-10~keV \citep{Gotthelf2011}, so the ratio between $\eta_{\text{R}}$ and $\eta_{\text{X}}$ is $\sim$10$^{-8}$. Both $\eta_{\text{R}}$ and $\eta_{\text{R}}$/$\eta_{\text{X}}$ are approximately 2 orders of magnitude lower than for pulsars of high $\dot{E}$ for which both radio and X-ray emission is seen \citep{Szary2014}, which hints at the possibility that the radio emission could be detected, but is instead beaming away.

We also targeted the CCO associated with the supernova remnant G296.5+10.0, PSR\,J1210$-$5226/1E\,1207.4$-$5209, which has been found to emit at X-ray wavelengths with a period of 424~ms \citep{Zavlin2000}. PSR J1210$-$5226 is one of only three CCOs to have a measured rotation rate. We see no evidence of radio pulsations above a flux density of 21~\uJy. The depth of this limit suggests that the CCO is radio-quiet at the time of the observation. The pulsed X-rays likely come from thermal emission from hotspots in the polar cap region, incidentally the same zone where radio emission would be generated for canonical pulsar radio emission \citep{Ruderman1975}. The opening angle of the X-ray emission is likely larger than the radio beam, so it would not be extraordinary to only detect the former. However, the presence of X-rays does not itself compel there to be radio emission. The very low surface magnetic field of around 10$^{11}$~G \citep{Halpern2015} that could be suppressing spin-powered radio emission may also be insufficient for heating to be provided from the impact of rotation-powered particles. Nevertheless, heating is still possible via internal cooling focused on the polar car by a strong crustal magnetic field \citep{Shabaltas2012}. Despite our deep upper limit we are still unable to rule out a geometrical explanation for a lack of radio emission. A pulsed radio detection or otherwise deeper upper limits are essential for understanding the physics governing multiwavelength emission from CCOs, and how this differs from other classes of neutron stars \citep{DeLuca2017}.

\section{Summary}\label{conclusion}
We have introduced and described a new large survey of supernova remnants, pulsar wind nebulae and unidentified TeV sources from the H.E.S.S. catalogue. The high sensitivity of MeerKAT has allowed upper limits on radio pulsations between 856-1712~MHz of around 30~\uJy. The 55 sources that have been targeted so far are composed of 32 supernova remnants, 17 candidate supernova remnants, two isolated pulsar wind nebulae and four unidentified TeV sources. As a result of our searches, two new pulsars have been discovered.
The positional alignment of the new pulsar \psrA, ~the candidate radio shell G022.045$-$0.028, the PWN G22.0+0.0, a characteristic age of 56~kyr and the pulsar being energetic enough to power the plerionic X-ray emission confirms within reasonable certainty that these are associated.

\psrB~is a faint, slow pulsar that we have discovered close to the centre of the shell-type SNR G15.9+0.2. Given the distance from DM, the slow rotation period and the presence of CCO J181852.0$-$150213, this pulsar is likely not associated with the remnant. We do not anticipate an evolutionary connection between the pulsar and the progenitor of the CCO, despite their close proximity on the sky. We do not detect radio pulsations from the CCO. We were unable to detect radio pulsations from five other CCOs and three candidate CCOs.
Both pulsar discoveries continue to be timed with MeerKAT and the Parkes radio telescope, and the search for new pulsars is ongoing. Once completed, we intend to use the statistics of the survey to perform a robust analysis by way of a population synthesis of young pulsars born from supernovae. With this, we aim to explain the factors that dictate the detectability of this population, such as beaming fractions, the influence of the interstellar medium, the Galactic neutron star birth rate and the luminosity distribution of radio pulsars.

\section*{Acknowledgements}
The MeerKAT telescope is operated by the South African Radio Astronomy Observatory (SARAO), which is a facility of the National Research Foundation, itself an agency of the Department of Science and Innovation. All the authors thank the staff at SARAO for scheduling the MeerKAT observations presented here. TRAPUM observations used the FBFUSE and APSUSE computing clusters for data acquisition, storage and analysis. These instruments were designed, funded and installed by the Max-Planck Institut f{\"u}r Radioastronomie (MPIfR) and the Max-Planck-Gesellschaft. PTUSE was developed with support from the Australian SKA Office and Swinburne University of Technology. The Parkes Radio Observatory is a part of the Australia Telescope National Facility (ATNF), which funded by the Government of Australia and administered by the Commonwealth Scientific and Industrial Research Organisation (CSIRO) national science agency. We acknowledge the Wiradjuri people as the traditional owners of the Parkes Observatory site. This work has made use of data obtained from the \textit{Suzaku} X-ray telescope, a collaborative mission between the space agencies of Japan (JAXA) and the USA (NASA).
JDT acknowledges funding from the United Kingdom's Research and Innovation Science and Technology Facilities Council (STFC) Doctoral Training Partnership, project code 2659479.
For the purpose of open access, the author has applied a Creative Commons Attribution (CC BY) licence to any Author Accepted Manuscript version arising.
We thank Dr.~W.~Reich (MPIfR) for help with compiling a list of targets.
We thank Dr.~I~Pastor~Marazeula and Dr.~A.~Basu (Jodrell Bank Centre for Astrophysics, The University of Manchester, UK) for their help with analysis of the scattering timescale of \psrA~and for help with estimating glitch activities, respectively.
We would like to thank the reviewer for their helpful comments and recommendations for improving this publication.
This work used version 1.69 of the ATNF Pulsar Catalogue and the December 2022 version of the Galactic SNR Catalogue by Green D. A, Cavendish Laboratory, Cambridge, United Kingdom.
SAOImage DS9 was used for image analysis. We also made use of APLpy, an open-source plotting package for Python \citep{aplpy2012, aplpy2019} and Astropy \citep{astropy2018}, a community-developed core Python package and an ecosystem of tools and resources for astronomy.
This research has made use of the SIMBAD database, operated at CDS, Strasbourg, France.
Radio images from the MAGPIS New 20cm GPS were retrieved from \href{https://third.ucllnl.org/gps/}{https://third.ucllnl.org/gps/}.

\section*{Data Availability}

Data that are not available through the public archive of the South African Radio Astronomy Observatory, and all source code, will be shared on reasonable request to the corresponding author. The project code for the TRAPUM Science Working Group is SCI-20180923-MK-03.

\bibliographystyle{mnras}
\bibliography{main}

\bsp	
\label{lastpage}
\end{document}